\documentclass[fleqn,10pt]{wlscirep}
\usepackage[utf8]{inputenc}
\usepackage[T1]{fontenc}
\usepackage{multirow}
\usepackage{amsmath}
\usepackage{centernot}
\usepackage{mathtools}
\usepackage{listings}
\title{Five rules for friendly rivalry in direct reciprocity}

\author[1]{Yohsuke Murase}
\author[2,*]{Seung Ki Baek}
\affil[1]{RIKEN Center for Computational Science, Kobe, Hyogo 650-0047, Japan}
\affil[2]{Department of Physics, Pukyong National University, Busan 48513, Korea}

\affil[*]{seungki@pknu.ac.kr}

\begin{abstract}
Direct reciprocity is one of the key mechanisms accounting for cooperation in
our social life. According to recent understanding, most of classical strategies
for direct reciprocity fall into one of two classes, `partners' or `rivals'. A
`partner' is a generous strategy achieving mutual cooperation, and a `rival'
never lets the co-player become better off. They have different working
conditions: For example, partners show good performance in a large population,
whereas rivals do in head-to-head matches. By means of exhaustive enumeration,
we demonstrate the existence of strategies that act as both partners and rivals.
Among them, we focus on a human-interpretable strategy, named `CAPRI' after its
five characteristic ingredients, i.e., cooperate, accept, punish, recover, and
defect otherwise. Our evolutionary simulation shows excellent performance of
CAPRI in a broad range of environmental conditions.
\end{abstract}
\begin{document}

\flushbottom
\maketitle
\thispagestyle{empty}

\section*{Introduction}
Theory of repeated games is one of the most fundamental mathematical
frameworks that has long been studied for understanding how and why
cooperation emerges in human and biological communities. Even when cooperation
cannot be a solution of a one-shot game, repetition can enforce cooperation
between the players by taking into account the possibility of future encounters.
A spectacular example is the prisoner's dilemma (PD) game: It describes a social
dilemma between two players, say, Alice and Bob, in which each player has two
options `cooperation' ($c$) and `defection' ($d$). The payoff matrix for the PD
game is defined as follows:
\begin{equation}
\left(
\begin{array}{cc|cc}
 &  & \multicolumn{2}{c}{\text{Bob}}\\
 &  & c & d\\\hline
\multirow{2}{*}{Alice} & c & (R,R) & (S,T) \\
                       & d & (T,S) & (P,P) \\
\end{array}
\right),
\label{eq:payoffs}
\end{equation}
where each entry shows (Alice's payoff, Bob's payoff) with $T>R>P>S$ and $2R >
T+S$. If the game is played once, mutual defection is the only equilibrium
because Alice maximizes her payoff by defecting no matter what Bob does.
However, if the game is repeated with sufficiently high probability, cooperation
becomes a feasible solution because the players have a strategic option that
they can reward cooperators by cooperating and/or they can punish defectors by
defecting in subsequent rounds (see, e.g., Table~\ref{tab:strategies}).
This is known as direct reciprocity, one of the
most well-known mechanisms for the evolution of
cooperation~\cite{nowak2006five}.

Through a series of studies, recent understanding of direct reciprocity proposes
that most of well-known strategies act either as partners or as
rivals~\cite{hilbe2015partners,hilbe2018partners}. Partner strategies are also
called `good strategies'~\cite{akin2015you,akin2016iterated}, and rival
strategies have been described as `unbeatable'~\cite{duersch2012unbeatable},
`competitive'~\cite{hilbe2015partners}, or
`defensible'~\cite{yi2017combination,murase2018seven}. Derived from our
everyday language, the `partner' and `rival' are defined as follows. As a
partner, Alice aims at sharing the mutual cooperation payoff $R$ with her
co-player Bob. However, when Bob defects from cooperation, Alice will punish Bob
so that his payoff becomes less than $R$. In other words, for Alice's strategy
to be a partner, we need the following two conditions: First, $\pi_{A} = \pi_{B}
= R$ when Bob applies the same strategy as Alice's, where $\pi_{A}$ and
$\pi_{B}$ represent the long-term average payoffs of Alice and Bob,
respectively. Second, when $\pi_{A}$ is less than $R$ because of Bob's
defection from mutual cooperation, $\pi_{B}$ must also be smaller than $R$,
whatever Bob takes as his strategy. It means that one of the best responses
against a partner strategy is choosing the same partner strategy so that they
form a Nash equilibrium. If a player uses a rival strategy, on the other hand,
the player aims at a payoff higher than or equal to the co-player's regardless
of the co-player's strategy. Thus, as long as Alice is a rival, it is guaranteed
that $\pi_{A} \geq \pi_{B}$. Note that these two definitions impose no
restriction on Bob's strategy, which means that the inequalities are unaffected
even if Bob remembers arbitrarily many previous rounds.

Which of these two traits is favoured by selection depends on environmental
conditions, such as the population size $N$ and the elementary payoffs $R$, $T$,
$S$, and $P$. For instance, a large population tends to adopt partner strategies
when $R$ is high enough. A natural question would be on the possibility that a
single strategy is \emph{both} a partner and a rival simultaneously: The point
is not to gain an extortionate payoff from the co-player in the sense of the
zero-determinant (ZD) strategies~\cite{press2012iterated} but to provide an
incentive to form mutual cooperation.
Let us call
such a strategy a `friendly rival' hereafter. Tit-for-tat (TFT) or Trigger
strategies can be friendly rivals in an ideal condition that the players are
free from implementation error due to ``trembling hands''. However, this is not
the case in a more realistic situation in which actions can be misimplemented
with probability $e > 0$. Here, the apparent contradiction between the notions
of a partner and a rival is seen as the most acute form. That is, Alice must
forgive Bob's erroneous defection to be a partner \emph{and} punish his
malicious defection to be a rival, without knowing Bob's intention. This is the
crux of the matter in relationships.

In this work, by means of massive supercomputing, we show that a tiny fraction
of friendly rival strategies exist among deterministic memory-three strategies
for the iterated PD game without future discounting. Differently from earlier
studies~\cite{press2012iterated,hilbe2013evolution,hilbe2014cooperation,hilbe2013adaptive,stewart2013extortion,stewart2014collapse,szolnoki2014defection,szolnoki2014evolution,mcavoy2016autocratic},
our strategies are deterministic ones, which makes each of them easy to
implement as a behavioural guideline as well as a public policy without
any randomization
device~\cite{dror1983public}. In particular, we focus on one of the friendly
rivals, named CAPRI, because it can be described in plain language, which
implies great potential importance in understanding and guiding human behaviour.
We also argue that our friendly rivals exhibit evolutionary
robustness~\cite{stewart2013extortion} for any population size and for any
benefit-to-cost ratio. This property is demonstrated by evolutionary simulation
in which CAPRI overwhelms other strategies under a variety of environmental
parameters.

\begin{table}%[htbp]
\centering
\caption{Description of well-known strategies in the iterated PD game. Whenever
possible, each strategy is represented as a tuple of five probabilities, i.e.,
$(p_0, p_{R}, p_{S}, p_{T}, p_{P})$, where $p_0$ means the probability to
cooperate in the first round, and $p_{\beta}$ means the probability to cooperate
after obtaining payoff $\beta$ in the previous round (see
Eq.~\ref{eq:payoffs}). Here, a zero-determinant (ZD) strategy has a positive
parameter $\phi$, and its other parameter $\eta$ lies in the unit
interval~\cite{press2012iterated,stewart2013extortion,hilbe2014extortion}.
}\label{tab:strategies}
\begin{tabular}{cc}
strategy & description \\
\midrule
AllC & $(1,1,1,1,1)$\\
AllD & $(0,0,0,0,0)$\\
Tit-for-tat (TFT) & $(1,1,0,1,0)$\\
Generous TFT & $(1,1,q,1,q)$ with $0<q<1$\\
Tit-for-two-tats (TF2T) & Defect if the co-player defected in the previous two
rounds.\\
Win-Stay-Lose-Shift (WSLS) & $(1,1,0,0,1)$\\
generous ZD
& $(1, 1, 1-\phi[(1-\eta)(S-R)+T-S],
\phi[(1-\eta)(R-T)+T-S], \phi(1-\eta)(R-P))$ \\
extortionate ZD
& $(0, 1-\phi(1-\eta)(R-P), 1-\phi[(1-\eta)(S-P)+T-S],
\phi[(1-\eta)(P-T)+T-S], 0)$ \\
Trigger & Defect if defection has ever been observed.\\
\bottomrule
\end{tabular}
\end{table}

\section*{Method}
%\subsection*{Ethics statement}
%No human experiments were conducted in this study.
%
%\subsection*{Algorithm}
Despite the fundamental importance of memory in direct reciprocity,
combinatorial explosion has been a major obstacle in understanding the memory
effects on cooperation: Let us consider deterministic strategies with memory
length $m$, which means that each of them chooses an action between $c$ and $d$
as a function of the $m$ previous rounds. The number of such memory-$m$
strategies expands as $N=2^{2^{2m}}$, which means $N_{m=1} = 16$, $N_{m=2} =
65536$, and $N_{m=3} \approx 1.84\times 10^{19}$. The number of combinations of
these strategies grows even more drastically, which renders typical evolutionary
simulation incapable of exploring the full strategy space. Here, we take an
axiomatic approach~\cite{yi2017combination,hilbe2017memory,murase2018seven} to
find friendly
rivals. That is, we search for strategies that satisfy certain predetermined
criteria, and the computation time for checking those criteria scales as $O(N)$
instead of $O(N^2)$ or greater.

\begin{figure}
\begin{center}
\includegraphics[width=\columnwidth]{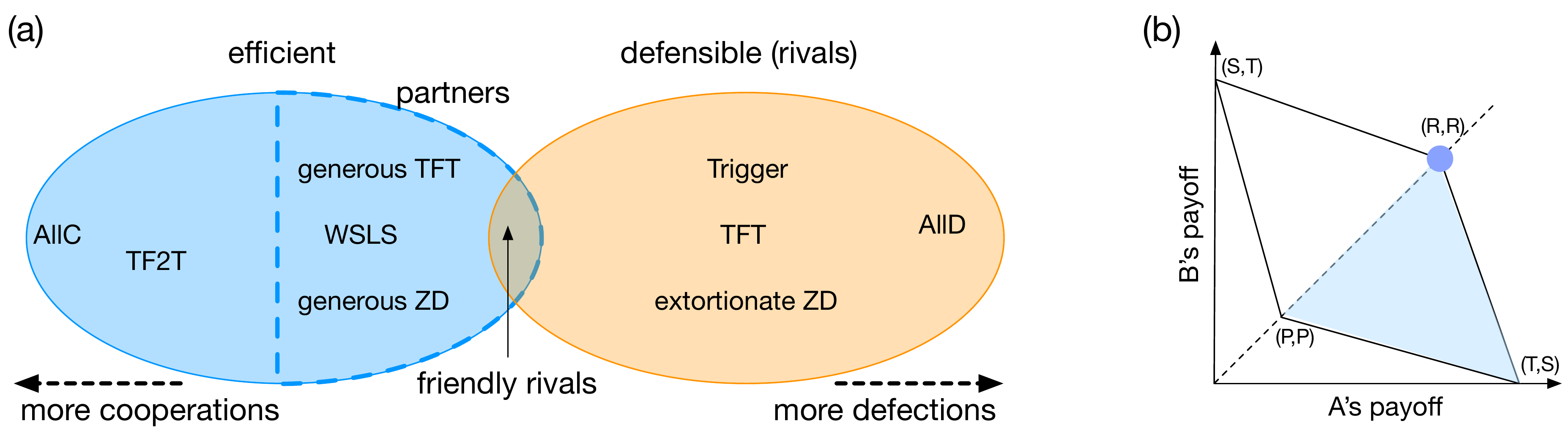}
\end{center}
\caption{ (Left) A schematic diagram of the strategy space. Strategies that tend
to cooperate (defect) are shown on the left (right). The blue ellipse represents
a set of efficient strategies, which are cooperative to sustain mutual
cooperation, and its subset of partner strategies is denoted by the dashed blue
curve.
On the other hand, the red ellipse represents a set of defensible
strategies, which often defect to defend themselves from malicious co-players.
In general, their intersection is small. When $m = 2$, for instance, the sizes
of efficient and defensible sets are $7639$ and $2144$, respectively, whereas
the intersection contains only eight strategies. (Right) The diamond depicts the
region of possible average payoffs for Alice and Bob. The blue triangle shows
the feasibility region when Alice uses a defensible strategy. If Alice and Bob
both use the same strategy satisfying efficiency, they will reach $(R,R)$ (the
blue dot).
}
\label{fig:strategy_space}
\end{figure}

More specifically, we begin with the following two
criteria~\cite{yi2017combination,murase2018seven}:
\begin{enumerate}
    \item Efficiency:
    Mutual cooperation is achieved with probability one as error probability $e
    \to 0^+$, if both Alice and Bob use this strategy.
    \item Defensibility:
    If Alice uses this strategy, she will never be outperformed by Bob when
    $e=0$ regardless of initial actions. This is a sufficient condition for
    being a rival, i.e., $\lim_{e\to 0^+}(\pi_{A} - \pi_{B}) \geq 0$.
\end{enumerate}
The efficiency criterion requires a strategy to establish cooperation in the
presence of small $e$ when both the players adopt this strategy. This criterion
is satisfied by many generous strategies such as unconditional cooperation
(AllC), generous TFT (GTFT), Win-Stay-Lose-Shift (WSLS) and Tit-for-two-tats
(TF2T). Partner strategies constitute a sub-class of efficient ones by limiting
the co-player's payoff to be less than or equal to $R$ regardless of the
co-player's payoff~\cite{akin2016iterated,hilbe2015partners,hilbe2018partners}.
On the other hand, a defensible strategy must ensure that the player's long-time
average payoff will be no less than that of the co-player who may use \emph{any}
possible strategy, and this idea is equivalent to the notion of a `rival
strategy'~\cite{hilbe2015partners,hilbe2018partners}. Defensible strategies
include unconditional defection (AllD), Trigger,
TFT, and extortionate ZD strategies.
Figure~\ref{fig:strategy_space}a schematically shows how these two criteria
narrow down the list of strategies to consider. The overlap of efficient and
defensible strategies means a set of friendly rivals because it is a subset of
partner strategies. It assigns the most strict limitation on the co-player's
payoff among the partner strategies as shown in
Fig.~\ref{fig:strategy_space}b. Indeed, the overlap region between these two
criteria is extremely tiny: It is pure impossibility for $m=1$, and we find only
$8$ strategies out of $N=65536$ for $m=2$.

To further narrow down the list of strategies, we impose the third
criterion~\cite{yi2017combination,murase2018seven}:
\begin{enumerate}
\setcounter{enumi}{2}
    \item Distinguishability: The strategy has a strictly higher payoff than the
    co-player's when its strategy is AllC in the small-error limit, i.e.,
    $\lim_{e\to 0^{+}}(\pi_{A} - \pi_{\rm AllC}) > 0$.
\end{enumerate}
This requirement originates from evolutionary game theory: If this criterion is
violated, the number of AllC players may increase due to neutral drift, which
eventually makes the population vulnerable to invasion of defectors such as
AllD. We check these criteria for each strategy by representing it as a
graph and analysing its topological properties (see Supplementary Methods
at the end of this manuscript). If a strategy satisfies all those three
criteria, it will be called `successful'.

Among deterministic memory-two strategies, it is known that only four strategies
satisfy these three criteria~\cite{yi2017combination}. They have minor
differences from each other, and one of them is called TFT-ATFT, which is a
combination of TFT and anti-tit-for-tat (ATFT). It usually behaves as TFT, but
it takes the opposite moves after mistakenly defecting from mutual cooperation.
Similar analysis has been conducted for the three-person public-goods (PG) game:
At least $256$ successful strategies exist when $m=3$, whereas no such solution
exists when $m<3$~\cite{murase2018seven}. It has also been shown that a friendly
rival strategy must have $m \ge n$ for the general $n$-person PG game, although
such a strategy for $n > 3$ is yet to be found. These results suggest that a
novel class of strategies may appear as the memory length exceeds a certain
threshold.

For memory-three strategies, we have obtained an exhaustive list of successful
strategies by massive supercomputing (see Supplementary Methods at the end of
this manuscript).
The efficiency and defensibility criteria find $7,018,265,885,034$ friendly
rivals out of $N_{m=3} = 2^{64} \approx 1.84 \times 10^{19}$ strategies. If the
distinguishability criterion is additionally imposed, $4,261,844,305,281$
strategies are found. There are four actions commonly prescribed by all these
successful strategies: Let $A_t$ and $B_t$ denote Alice's and Bob's actions at
time $t$, respectively. When their memory states are
$(A_{t-3}A_{t-2}A_{t-1},B_{t-3}B_{t-2}B_{t-1}) = (ccc,ccc)$, $(ccc,ddd)$,
$(cdd,ddd)$, and $(ddd,ddd)$, all the successful strategies tell Alice to choose
$c$, $d$, $d$, and $d$, respectively. The first one is absolutely required to
maintain mutual cooperation. The latter three are needed to satisfy the
defensibility criterion: If $c$ was prescribed at any of these states, Alice
would be exploited by Bob's continual defection.

\begin{table}%[htbp]
\centering
\caption{
Recovery paths to mutual cooperation for the memory-three successful strategies.
Only the most common five patterns are shown in this table.
The upper and lower rows represent the sequences of actions taken by Alice and Bob, respectively, when Bob defected from mutual cooperation by error.
The right column shows the number of strategies having each pattern, as well as its fraction with respect to the total number of successful strategies.
}\label{tab:recovery_pattern}
\begin{tabular}{lr}
action sequence & \# of strategies \\
\midrule
  $\begin{matrix}\dots & c & c & d & c & \dots \\ \dots & c & d & c & c & \dots \end{matrix}$
& \begin{tabular}{r} 905,772,524,235 \\ (21.3\%) \end{tabular} \\
  $\begin{matrix}\dots & c & c & d & d & c & \dots \\ \dots & c & d & c & d & c & \dots \end{matrix}$
& \begin{tabular}{r} 522,061,013,252 \\ (12.2\%) \end{tabular} \\
  $\begin{matrix}\dots & c & c & d & d & c & \dots \\ \dots & c & d & d & c & c & \dots \end{matrix}$
& \begin{tabular}{r} 437,671,509,356 \\ (10.3\%) \end{tabular} \\
  $\begin{matrix}\dots & c & c & d & c & d & c & \dots \\ \dots & c & d & c & c & d & c & \dots \end{matrix}$
& \begin{tabular}{r} 409,458,612,318 \\ (9.6\%) \end{tabular} \\
  $\begin{matrix}\dots & c & c & d & c & d & d & c & \dots \\ \dots & c & d & c & c & d & d & c & \dots \end{matrix}$
& \begin{tabular}{r} 227,113,898,468 \\ (5.3\%) \end{tabular} \\
  $\begin{matrix}\dots & c & c & d & d & d & c & \dots \\ \dots & c & d & c & c & d & c & \dots \end{matrix}$
& \begin{tabular}{r} 184,052,002,852 \\ (4.3\%) \end{tabular} \\
\bottomrule
\end{tabular}
\end{table}

Except for these four prescriptions, we see a wide variety of patterns. For
example, let us assume that both Alice and Bob adopt one of these strategies.
When Bob defects by error, they must follow a recovery path from state
$(ccc,ccd)$ to $(ccc,ccc)$. We find $839$ different patterns from our successful
strategies (Table~\ref{tab:recovery_pattern}). The most common one is also the
shortest, in which only two time steps are needed to recover mutual cooperation.
It cannot be shorter because Alice must defect at least once to assure
defensibility. It is even shorter than that of TFT-ATFT, which is identical to
the third entry of Table~\ref{tab:recovery_pattern}. This finding disproves a
speculation that friendly rivals are limited to
variants of TFT even if $m>2$~\cite{yi2017combination}. This
shortest recovery path is possible only when $m \ge 3$, indicating a pivotal
role of memory length in direct reciprocity.

\section*{Result}

\subsection*{CAPRI strategy}
The shortest recovery path in Table~\ref{tab:recovery_pattern} shows that Bob
can recover his own mistake simply by accepting Alice's punishment provided that
he has $m=3$. Among the strategies using this recovery pattern, we have
discovered a strategy which is easy to interpret, named `CAPRI', after the first
letters of its five constitutive rules listed below:
\begin{enumerate}
    \item Cooperate at mutual cooperation. This rule prescribes $c$ at
    $(ccc,ccc)$.
%    \begin{itemize}
%        \item $(ccc,ccc) \to c$
%    \end{itemize}
    \item Accept punishment when you mistakenly defected from mutual
    cooperation. This rule prescribes $c$ at $(ccd,ccc)$, $(cdc,ccd)$,
    $(dcc,cdc)$, and $(ccc,dcc)$.
%    \begin{itemize}
%        \item $(ccd,ccc) \to c$
%        \item $(cdc,ccd) \to c$
%        \item $(dcc,cdc) \to c$
%        \item $(ccc,dcc) \to c$
%    \end{itemize}
    \item Punish your co-player by defecting once when he defected from mutual
    cooperation. This rule prescribes $d$ at $(ccc,ccd)$, and then $c$ at
    $(ccd,cdc)$, $(cdc,dcc)$, and $(dcc,ccc)$.
%    \begin{itemize}
%        \item $(ccc,ccd) \to d$
%        \item $(ccd,cdc) \to c$
%        \item $(cdc,dcc) \to c$
%        \item $(dcc,ccc) \to c$
%    \end{itemize}
    \item Recover cooperation when you or your co-player cooperated at mutual
    defection. This rule prescribes $c$ at $(ddd,ddc)$, $(ddc,dcc)$,
    $(dcc,ccc)$, $(ddc,ddd)$, $(dcc,ddc)$, $(ccc,dcc)$, $(ddc,ddc)$, and
    $(dcc,dcc)$.
%    \begin{itemize}
%        \item $(ddd,ddc) \to c$
%        \item $(ddc,dcc) \to c$
%        \item $(dcc,ccc) \to c$ (duplicate of rule 3)
%        \item $(ddc,ddd) \to c$
%        \item $(dcc,ddc) \to c$
%        \item $(ccc,dcc) \to c$ (duplicate of rule 2)
%        \item $(ddc,ddc) \to c$
%        \item $(dcc,dcc) \to c$
%    \end{itemize}
    \item In all the other cases, defect.
\end{enumerate}

\begin{figure}
\begin{center}
\includegraphics[width=0.8\columnwidth]{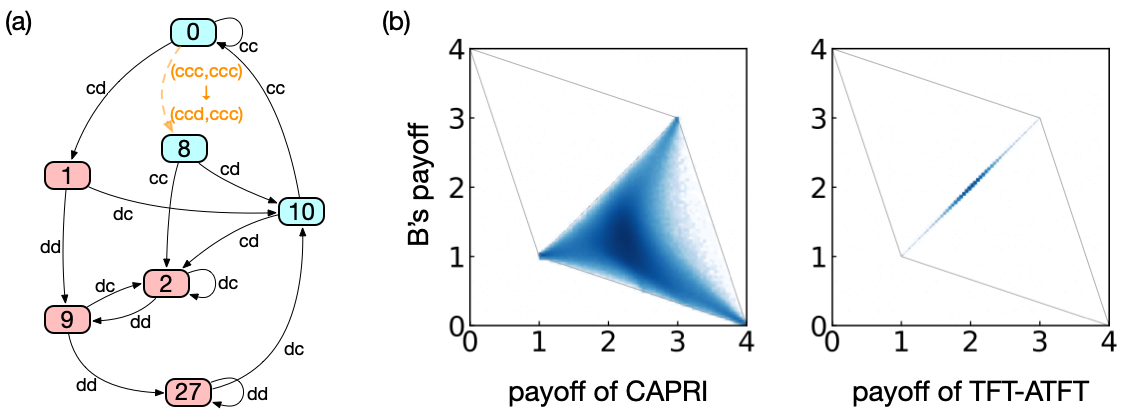}
\end{center}
\caption{
(a) Automaton representation of CAPRI. Its prescribed actions are denoted by the
node colours (blue for $c$ and red for $d$). The labels on the edges indicate the
players' actions. The transition caused by erroneous defection at mutual
cooperation (`0') is depicted by an orange dashed arrow. (b) Distribution of
payoffs when Alice's strategy is CAPRI (left) or TFT-ATFT (right), whereas Bob
adopts one of probabilistic memory-three strategies uniformly at random. The
elementary payoffs are $(R,T,S,P)=(3,4,0,1)$.
}
\label{fig:ETS_automaton}
\end{figure}

The first rule is clearly needed for efficiency. In addition, mutual cooperation
must be robust against one-bit error, i.e., occurring with probability of
$O(e)$, when both Alice and Bob use this strategy. This property is provided by
the second and the third rules. In addition, for this strategy to be efficient,
the players must be able to escape from mutual defection through one-bit error
so that the stationary probability distribution does not accumulate at mutual
defection, which is handled by the fourth rule. Note that these four rules for
efficiency do not necessarily violate defensibility when $m>2$, as we have
already seen in Table~\ref{tab:recovery_pattern}. Actually, due to the fifth
rule, both efficiency and defensibility are satisfied by CAPRI. The action table
and its minimized automaton representation~\cite{murase2019automata} are given
in Table~\ref{tab:ETS} and Fig.~\ref{fig:ETS_automaton}a, respectively. The
self-loop via $dc$ at state `2' in Fig.~\ref{fig:ETS_automaton}a proves that
this strategy also satisfies distinguishability.

CAPRI requires $m=3$ because otherwise it violates defensibility: If CAPRI were
a memory-two strategy, $(cd,dc)\to c$ and $(dc,cd)\to c$ must be prescribed to
recover from error. However, with these prescriptions, Bob can repeatedly
exploit Alice by using the following sequence:
\begin{equation}
    \begin{matrix}\dots & c & c & d & c & c & \dots \\
                  \dots & c & d & c & d & c & \dots.
    \end{matrix}
\end{equation}
TFT-ATFT and its variants are the only friendly rivals when $m<3$.
Compared with TFT-ATFT, CAPRI is closer to Grim trigger (GT) rather than to TFT.
Alice keeps cooperating as long as Bob cooperates, but she switches to
defection, as prescribed by the fifth rule, when Bob does not conform to her
expectation. Due to the similarity of CAPRI to GT, it also outperforms a wider
spectrum of strategies than TFT-ATFT. Figure~\ref{fig:ETS_automaton}b shows the
distribution of payoffs of the two players when Alice's strategy is CAPRI, and
Bob's strategy is sampled from the $64$-dimensional unit hypercube of
memory-three probabilistic strategies. Alice's payoff is strictly higher than
Bob's in most of the samples. On the other hand, when Alice uses TFT-ATFT,
payoffs are mostly sampled on the diagonal because it is based on TFT, which
equalizes the players' payoffs.
However, we also note that CAPRI is significantly different from GT in two ways.
First, CAPRI is error-tolerant: Even when Bob makes a mistake, Alice is ready to
recover cooperation after Bob accepts punishment, as described in the second and
the third rules. Second, whereas GT is characterized by its irreversibility,
CAPRI lets the players escape from mutual defection according to the fourth
rule.

\begin{table}%[htb]
\centering
\caption{
Action table of CAPRI. The superscript on the upper left corner of each element
indicates which rule is involved.
%cdddcdddcdcddddddcddcdddddddddddcdcdcdcdddddddddddddcdccddddddcd
}\label{tab:ETS}
\begin{tabular}{ccccccccc}
      & \multicolumn{8}{c}{$B_{t-3}B_{t-2}B_{t-1}$} \\
$A_{t-3}A_{t-2}A_{t-1}$ & $ccc$ & $ccd$ & $cdc$ & $cdd$ & $dcc$ & $dcd$ & $ddc$ & $ddd$\\
\midrule
$ccc$ & $^1c$ & $^3d$ & $d$ & $d$ & $^{2,4}c$ & $d$ & $d$ & $d$ \\
$ccd$ & $^2c$ & $d$ & $^3c$ & $d$ & $d$ & $d$ & $d$ & $d$ \\
$cdc$ & $d$ & $^2c$ & $d$ & $d$ & $^3c$ & $d$ & $d$ & $d$ \\
$cdd$ & $d$ & $d$ & $d$ & $d$ & $d$ & $d$ & $d$ & $d$ \\
$dcc$ & $^{3,4}c$ & $d$ & $^2c$ & $d$ & $^4c$ & $d$ & $^4c$ & $d$ \\
$dcd$ & $d$ & $d$ & $d$ & $d$ & $d$ & $d$ & $d$ & $d$ \\
$ddc$ & $d$ & $d$ & $d$ & $d$ & $^4c$ & $d$ & $^4c$ & $^4c$ \\
$ddd$ & $d$ & $d$ & $d$ & $d$ & $d$ & $d$ & $^4c$ & $d$ \\
\bottomrule
\end{tabular}
\end{table}

\subsection*{Evolutionary simulation}

\begin{figure*}
\begin{center}
\includegraphics[width=\textwidth]{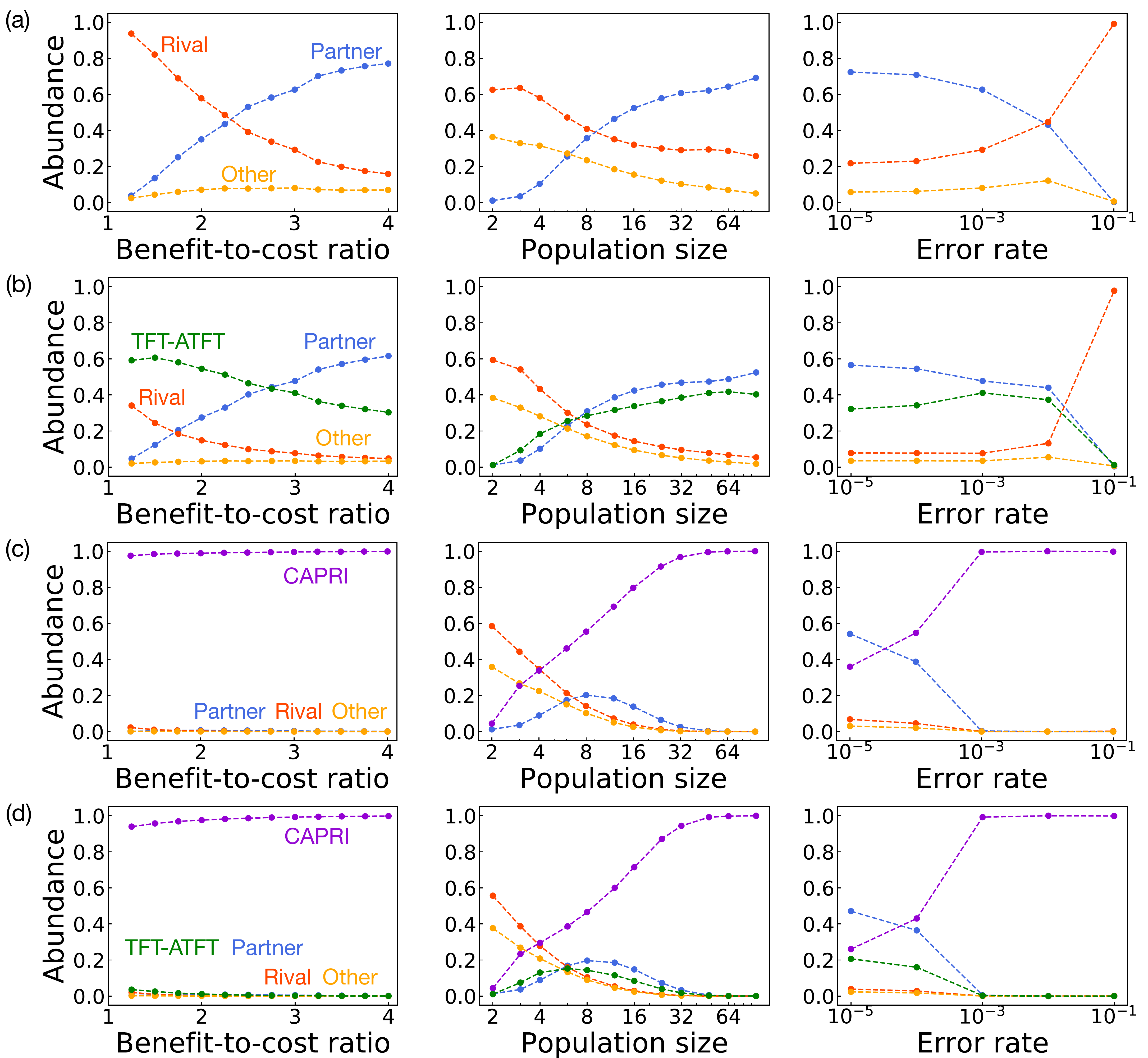}
\end{center}
\caption{
(a) Abundance of memory-one partners, rivals, and the other strategies. We
consider a simplified version of the PD game, parametrized by $b$,
the benefit to the co-player when a player cooperates with a unit cost.
In terms of the elementary payoffs, this corresponds to $R = b-1$, $T=b$,
$S=-1$, and $P=0$. The Moran process is simulated with selection strength
$\sigma$ in a population of size $N$, where the product $N \sigma$ is fixed as
$10$.
Three parameters (benefit-cost-ratio $b$, population size $N$, and error
rate $e$) are varied one by one~\cite{hilbe2018partners}. Their default values
are $b=3$, $N=50$, and $e=10^{-3}$, unless otherwise stated. We also show the
simulation results with (b) TFT-ATFT, (c) CAPRI, and (d) both TFT-ATFT and
CAPRI, introduced with probability $\mu = 0.01$. These are average results over
10 Monte-Carlo runs.
}
\label{fig:evo_CAPRI}
\end{figure*}

Although defensibility assures that the player is never outperformed by the
co-player, it does not necessarily guarantee success in evolutionary games,
where everyone is pitted against every other in the population. For example,
extortionate ZD strategies perform poorly in an evolutionary
game~\cite{stewart2013extortion,hilbe2013adaptive,adami2013evolutionary}. In
this section, we will check the performance of CAPRI in the evolutionary
context.

When we consider performance of a strategy in an evolving population, the most
famous measure of assessment is evolutionary stability
(ES)~\cite{smith1982evolution}. Although conceptually useful, ES is too
strong a condition, requiring that when a sufficient majority of population
members apply the strategy, every other strategy is at a selective disadvantage.
Evolutionary robustness has thus been introduced as
a more practical notion of stability~\cite{stewart2013extortion}: A
strategy is called evolutionary robust if no other strategy has fixation
probability greater than $1/N$, which is the fixation probability of a neutral
mutant. In other words, an evolutionary robust strategy cannot be selectively
replaced by any mutant strategy. Evolutionary robustness of a strategy depends
on the population size: Partner strategies have this property when $N$ is large
enough, whereas for rival strategies, it is when $N$ is
small~\cite{stewart2013extortion}. Friendly rivals have the virtue of both:
They keep evolutionary robustness regardless of $N$, as will be shown below.

As in the standard stochastic model~\cite{imhof2010stochastic}, let
us consider a well-mixed population of size $N$ in which selection follows an
imitation process. At each discrete time step, a pair of players are chosen at
random, and we will call their strategies $X$ and $Y$, respectively. The
probability for one of the players to replace her strategy $X$ with $Y$ is given
as follows:
\begin{equation}
    f_{x \to y} = \frac{1}{1 + \exp\left[\sigma \left(s_x - s_y\right) \right]},
\end{equation}
where $s_x$ and $s_y$ denote the average payoffs of $X$ and $Y$ against the
entire population, respectively, and $\sigma$ is a parameter which denotes the
strength of selection. In population dynamics, we assume that the mutation rate
$\mu$ is low enough: That is, when a mutant strategy $X$ appears in a resident
population of $Y$, no other mutant will be introduced until $X$ reaches fixation
or goes extinct. The dynamics is formulated as a Moran process, under which the
fixation probability of $X$ is given in a closed
form~\cite{stewart2013extortion}:
\begin{equation}
    \rho = \frac{1}{\sum_{i=0}^{N-1}\prod_{j=1}^{i} e^{\sigma\left[(N-j-1)s_{yy} + js_{yx} - (N-j)s_{xy} - (j-1)s_{xx} \right] }},
\end{equation}
where $s_{xy}$ denotes the long-term payoff of player $X$ against player $Y$.
Using Jensen's inequality, we see that
%\begin{eqnarray}
%    \frac{1}{\rho} &=& \sum_{i=0}^{N-1} e^{ \sigma i \left[ (2N-i-3)s_{yy}  +
%    (i+1)s_{yx} - (2N-i-1)s_{xy} -(i-1)s_{xx} \right] /2 } \\
%    &\geq& N e^{ \sigma (N-1) \left[ (N-2)(s_{yy}-s_{xx}) + (N+1)(s_{yx}-s_{xy})
%    + (N-2)(s_{yy} - s_{xy}) \right] /6 }.
%\label{eq:evo_rho}
%\end{eqnarray}
\begin{equation}
\frac{1}{\rho} = \sum_{i=0}^{N-1} e^{ \sigma i \left[ (2N-i-3)s_{yy}  +
(i+1)s_{yx} - (2N-i-1)s_{xy} -(i-1)s_{xx} \right] /2 }
\end{equation}
\begin{equation}
\geq N e^{ \sigma (N-1) \left[ (N-2)(s_{yy}-s_{xx}) + (N+1)(s_{yx}-s_{xy}) +
(N-2)(s_{yy} - s_{xy}) \right] /6 }.
\end{equation}
When $Y$ is a partner strategy, it satisfies $s_{yy} \ge s_{xy}$ and $s_{yy} \ge s_{xx}$. When $Y$ is also a rival strategy, it has another inequality, $s_{yx}
\ge s_{xy}$. Therefore, the fixation probability of an arbitrary mutant $\rho
\leq 1/N$ regardless of $N$ and $\sigma$.

We have conducted evolutionary simulation to assess the performance of friendly
rivals. First, we run simulation without CAPRI and TFT-ATFT. This simulation
adopts the setting of a recent study~\cite{hilbe2018partners}
and serves as a baseline of
performance. A mutant strategy is restricted to reactive memory-one strategies,
according to which the player's action depends only on the co-player's last
action. The reactive strategies are characterized by a pair of probabilities
($p_{c}$,$p_{d}$), where $p_\alpha$ denotes the probability to cooperate when
the co-player's last move was $\alpha$. Rival strategies are represented by $p_d
= 0$, and partners are by $p_c = 1$ and $p_d < p_d^{\ast}$, where $p_d^{\ast}
\equiv \min\{1-(T-R)/(R-S),(R-P)/(T-P)\}$. Mutant strategies may be randomly
drawn from $[0,1] \times [0,1]$, but we have discretized the unit square in a
way that each $p_\alpha$ takes a value from $\{0, 1/10, 2/10, \dots, 9/10, 1\}$.
We have run the simulation until mutants are introduced $10^7$ times,
and measured how frequently partner or rival strategies are observed. As shown
in Fig.~\ref{fig:evo_CAPRI}a, evolutionary performance of strategies depends
on environmental
parameters~\cite{stewart2013extortion,stewart2014collapse,hilbe2018partners}.
Rival strategies have higher abundance when the benefit-to-cost ratio is low,
population size $N$ is small, and error rate $e$ is high. Otherwise, partner
strategies are favoured.

Let us now assume that a mutant can also take TFT-ATFT in addition to the
reactive memory-one strategies. Figure~\ref{fig:evo_CAPRI}b shows that
TFT-ATFT occupies significant fractions across a broad range of parameters. The
situation changes even more remarkably when CAPRI is introduced instead of
TFT-ATFT. As seen in Fig.~\ref{fig:evo_CAPRI}c, CAPRI overwhelms the other
strategies for almost the entire parameter ranges. The low abundance at $N=2$ or
$e=10^{-5}$ does not contradict with the evolutionary robustness of CAPRI
because it is still higher than the abundance of a neutral mutant.
Although the abundance of partners is higher than CAPRI when $e=10^{-5}$, the
reason is that it is an aggregate over many partner strategies. If we compare
each single strategy, CAPRI is still the most abundant one for the entire range
of $e$.
The qualitative picture remains the same even if we choose a different
value of $\sigma$, and CAPRI tends to be more favoured as $\sigma$ increases.
Furthermore,
by comparing Figs.~\ref{fig:evo_CAPRI}b and \ref{fig:evo_CAPRI}c, we see
that CAPRI shows better performance than TFT-ATFT. The evolutionary advantage of
CAPRI over TFT-ATFT is directly observed in Fig.~\ref{fig:evo_CAPRI}d, where
both CAPRI and TFT-ATFT are introduced into the population. As we have seen in
Fig.~\ref{fig:ETS_automaton}b, it tends to earn strictly higher payoffs
against various types of co-players, whereas TFT-ATFT, based on TFT, aims to
equalize the payoffs except when it encounters naive cooperators. This
observation shows a considerable amount of diversity even among evolutionary
robust strategies~\cite{stewart2016small}.

\section*{Summary}\label{sec:summary}
To summarize, we have investigated the possibility to act as both a partner and
a rival in the repeated PD game without future discounting. By thoroughly
exploring a huge number of strategies with $m=3$, we have found that it is
indeed possible in various ways. The resulting friendly rivalry directly implies
evolutionary robustness for any population size, benefit-to-cost ratio, and
selection strength. We observe its success even when $e$ is of a considerable
size (Fig.~\ref{fig:evo_CAPRI}). It is also worth noting that a friendly rival
can publicly announce its strategy because it is guaranteed not to be
outperformed regardless of the co-player's prior knowledge. Rather, it is
desirable that the strategy should be made public because the co-player can be
advised to adopt the same strategy by knowing it from the beginning to maximize
its payoff. The resulting mutual cooperation is a Nash equilibrium. The
deterministic nature offers additional advantages because the player can
implement the strategy without any randomization device. Moreover, even if
uncertainty exists in the cost and benefit of cooperation, a friendly rival
retains its power because it is independent of $(R,T,S,P)$. This is a distinct
feature compared to the ZD strategies, whose cooperation probabilities have to
be calculated from the elementary payoffs. Furthermore, the results are
independent of the specific payoff ordering $T>R>P>S$ of the PD. These are valid
as long as mutual cooperation is socially optimal ($R>P$ and $2R > T+S$) and
exploiting the other's cooperation pays better than being exploited ($T>S$).
This condition includes other well-known social dilemma, such as the snowdrift
game (with $T>R>S>P$) and the stag-hunt game (with $R>T>P>S$).

This work has focused on one of friendly rivals, named CAPRI. We speculate that
it is close to the optimal one in several respects: First, it recovers mutual
cooperation from erroneous defection in the shortest time. Second, it
outperforms a wide range of strategies. Furthermore, its simplicity is almost
unparalleled among friendly rivals discovered in this study. CAPRI is explained
by a handful of intuitively plausible rules, and such simplicity greatly
enhances its practical applicability because the required cognitive load will be
low when we humans play the
strategy~\cite{wedekind1996human,milinski1998working,hilbe2014cooperation}. It
is an interesting research question whether this statement can be verified
experimentally.

In particular, we would like to stress the importance of memory length in theory
and experiment, considering that much research attention has been paid to the
study of memory-one
strategies~\cite{stewart2013extortion,stewart2014collapse,akin2016iterated,hilbe2015partners,baek2008intelligent,hilbe2018indirect,ichinose2018zero}.
Besides the combinatorial explosion of strategic possibilities, one can argue
that a memory-one strategy, if properly designed, can unilaterally control the
co-player's payoff even when the co-player has longer
memory~\cite{press2012iterated}. It has also been shown that $m=1$ is enough
for evolutionary robustness against mutants with longer
memory~\cite{stewart2013extortion}. However, the payoff that a strategy receives
against itself may depend on its own memory
capacity\cite{stewart2013extortion,stewart2016small}, and this is the
reason that a friendly rival is feasible when $m>1$. We can gain some important
strategic insight only by moving beyond $m=1$.
Related to the above point, one of important open problems is how to design a
friendly-rival strategy for multi-player games. Little is known of the
relationship between a solution of an $n$-person and that of an $(n-1)$-person
game of the same kind. For example, it is known that TFT-ATFT for the PD
game~\cite{yi2017combination} is not directly applicable to the three-person PG
game~\cite{murase2018seven}. We nevertheless hope that the five rules of CAPRI
may be more easily generalized to the $n$-person game, considering that its
working mechanism seems more comprehensible than that of TFT-ATFT to the human
mind.

In a broader context, although `friendly rivalry' sounds self-contradictory, the
term captures a crucial aspect of social interaction when it goes in a
productive way: Rivalry is certainly ubiquitous between artists, sports teams,
firms, research groups, or neighbouring
countries~\cite{hogan2007behind,brandenburger2011co,kilduff2014driven,pike2018long}.
At the same time, they are subject to repeated interaction, whereby they
eventually become friends, colleagues, or business partners to each other. Our
finding suggests that such a seemingly unstable relationship can readily be
sustained just by following a few simple rules: Cooperate if everyone does,
accept punishment for your mistake, punish defection, recover cooperation if you
find a chance, but in all the other cases, just take care of yourself. These
seem to be the constituent elements for such a sophisticated compound of rivalry
and partnership.

%\bibliography{capri}

\section*{Acknowledgements}
The authors would like to thank C.~Hilbe for his careful reading and
insightful comments on the manuscript. Y.M. acknowledges support from MEXT as
``Exploratory Challenges on Post-K computer (Studies of multi-level
spatiotemporal simulation of socioeconomic phenomena)'' and from Japan Society
for the Promotion of Science (JSPS) (JSPS KAKENHI; Grant no. 18H03621). S.K.B.
acknowledges support by Basic Science Research Program through the National
Research Foundation of Korea (NRF) funded by the Ministry of Education
(NRF-2020R1I1A2071670).
This research used computational resources of the K computer provided by the
RIKEN Center for Computational Science through the HPCI System Research project
(Project ID:hp160264). OACIS was used for the simulations in this
study~\cite{murase2018open}.
We acknowledge the hospitality at APCTP where part of
this work was done.

\section*{Author contributions statement}
Y.M. designed the research, carried out
the computation, and analysed the results. S.K.B. verified the method. Y.M.
and S.K.B. wrote and reviewed the manuscript.

\section*{Additional information}

The authors declare no competing interests.
The source code for this
study is available at \url{https://github.com/yohm/sim_exhaustive_m3_PDgame}.

\newpage

\section*{Supplementary Methods}
\label{appendix:method}

We wish to examine the strategy space of $m=3$, but
it is impossible to enumerate all the memory-three strategies by a naive brute-force method even if we use a cutting-edge supercomputer because their total number
is as large as $2^{2^{nm}} = 2^{64} \approx 2 \times 10^{19}$.
To overcome this difficulty, we have developed graph-theoretic algorithms to judge defensibility, efficiency, and distinguishability.
In the following, we explain the algorithm in three steps: First, we present basic ideas to judge the three criteria for a single strategy.
Second, we show how this can be done for a set of strategies simultaneously.
Third, we apply these algorithms to enumerate all successful strategies comprehensively in the memory-three strategy space.
A C++ source code is available under an open-source license at \url{https://github.com/yohm/sim_exhaustive_m3_PDgame}.

\subsection*{Judging the criteria for a single strategy}\label{subsec:judge_strategy}

Let us consider two players $A$ and $B$ in the iterated PD game. Player $A$'s action at time $t$ is denoted as $A_t$, and $B_t$ is defined likewise. When $m=3$, we have $64$ different history profiles, $(A_{t-3}A_{t-2}A_{t-1},B_{t-3}B_{t-2}B_{t-1}) = (ccc,ccc)$, $(ccc,ccd)$, $(ccc,cdc)$, $\dots (ddd,ddd)$. These profiles can also be represented as $0$, $1$, $2$, $\dots 63$ in binary.
Consider a directed graph whose nodes represent the history profiles
and whose links represent transition among them as prescribed by $S_A$ and $S_B$, where $S_A$ and $S_B$ are the strategies of the players $A$ and $B$, respectively.
Such a graph will be called a transition graph in general.
Due to the deterministic property of $S_A$ and $S_B$, each node has one outgoing link in the absence of error, so the total number of links is also $64$. We will denote this graph as $g(S_A, S_B)$.

We may also consider another transition graph for the case where $B$'s actions are left undetermined whereas $A$'s strategy is $S$, namely $g(S, \ast)$.
Player $B$ may choose either $c$ or $d$, thus each node has two outgoing links.
This graph is useful in judging the defensibility of $S$: This criterion concerns relative payoff differences, which are made by either unilateral cooperation or unilateral defection.
Traversing every possible cycle in $g(S,\ast)$, therefore, we count the number of nodes with $(A_{t-1}, B_{t-1})=(c,d)$ and subtract it from the number of nodes with $(A_{t-1}, B_{t-1})=(d,c)$.
Working with integer counts is also numerically convenient.
If none of the cycles in $g(S,\ast)$ gives a negative value in this counting, we can say that the strategy $S$ satisfies the defensibility criterion.

A conventional way to judge the efficiency criterion of $S$ is to consider a transition graph of $S$ against itself, but with error probability $e$. Due to error, both the players can choose either $c$ or $d$, which means that each node of the transition graph has four outgoing links. One can check the corresponding stationary probability $\vec{\pi} = (\pi_0, \ldots, \pi_{63})^\intercal$, where $\intercal$ means transpose.
The strategy $S$ is efficient if $\pi_0$ converges to 100\%
as the error probability $e$ approaches zero from above.
The calculation of $\vec{\pi}$ can be done through linear algebraic calculation
with decreasing $e$ gradually~\cite{yi2017combination,murase2018seven}.

The above method takes into account the effects of error all at once.
However, we can devise a quicker way to judge the efficiency criterion \emph{topologically} with increasing the order of $e$ one by one, as we will explain now.
Let us begin with $g_0 = g(S, S)$ which does not take into account any error.
It is a directed graph, either connected or disconnected.
We write $i \rightarrow j$ if node $j$ is reachable from $i$ in $g_0$, and $i \not\rightarrow j$ otherwise.
When they are mutually reachable (unreachable), we write $i \leftrightarrow j$ ($i \not\leftrightarrow j$).
If a node has no outgoing links, it is called a sink.
This notion is extended to a strongly connected component (SCC) as well, that is, a SCC is also called a sink if it has no outgoing links.
If $\alpha$ is a SCC composed of nodes $\alpha_1, \alpha_2, \ldots, \alpha_s$, the stationary distribution over $\alpha$ is defined as $\pi_\alpha = \pi_{\alpha_1} + \pi_{\alpha_2} + \ldots + \pi_{\alpha_s}$.
If a sink is reachable from a node, we say that the node is in the basin of the sink, where the basin includes the sink itself.

If $i \rightarrow 0$ and $0 \not\rightarrow i$ for every node $i \neq 0$ in $g_0$, the node $0$ constitutes the unique sink of this graph: It is a sink, by definition. It is also unique because none of the other nodes is a sink. If this is the case, just by checking $g_0$ without considering error, we may conclude that the strategy $S$ under consideration satisfies the efficiency criterion. Although we have not taken into account error-induced transitions, this conclusion can be justified in two ways: First, the detailed-balance condition implies that $\pi_i/\pi_0 \lesssim O(e)$ for every $i \neq 0$ because $i$ can be accessed from the sink $0$ only by error. Or, we can explicitly construct the derivative of the principal eigenvector by using the fact that it is non-degenerate~\cite{van2007computation}, which implies that error-induced change in connectivity with a size of $e \ll 1$ perturbs the stationary distribution $\vec{\pi}$ by an amount of $O(e)$ at most. As $e\to 0^+$, therefore, $\pi_0$ will approach 100\%.
% By the same logic, we can also say that strategy $S$ cannot be efficient if $g_0$ has no sink.

However, if the above condition is not met, i.e., if multiple sinks coexist, it is impossible to judge the efficiency of $S$ from $g_0$. We then need to take error-induced transitions into consideration. For example, let us consider $g_0$ with two sinks $\alpha$ and $\beta$, together with their respective basins $\Omega_\alpha^{(0)}$ and $\Omega_\beta^{(0)}$. In other words, every node $i$ in this graph satisfies $i \to \alpha$ or $i \to \beta$.
We define $\rho_\alpha^{(1)}$ as a set of nodes that are reachable from $\alpha$ via a single error, and define $\rho_\beta^{(1)}$ likewise. Formally speaking, we define $\rho_\alpha^{(k)} \equiv \{i | \alpha \xrightarrow[k]{} i \}$, where the subscript $k$ below the arrow means that we are considering transitions that are mediated by $k$ errors at least. The simple reachability relation $\alpha \to i$ is equivalent to $\alpha \xrightarrow[0]{} i$.
We will assume that $\rho_\beta^{(1)}$ has an overlap with $\Omega_\alpha^{(0)}$ at node $j$, whereas $\rho_\alpha^{(1)}$ does not with $\Omega_\beta^{(0)}$. It means that $\beta \xrightarrow[1]{} j \to \alpha$ whereas $\alpha \centernot{\xrightarrow[1]{}} \beta$. Then, for every node $i$ in $\Omega_\beta^{(0)}$, we find the following path: $i \to \beta \xrightarrow[1]{} j \to \alpha$. Strictly speaking, $\alpha$ is no longer a sink at this level of description because $\alpha \xrightarrow[1]{} l \in \rho_\alpha^{(1)}$ by the definition of $\rho_\alpha^{(1)}$. However, we expect that such transitions should not alter the situation significantly because $\rho_\alpha^{(1)}$ is still a subset of $\Omega_\alpha^{(0)}$, meaning that the dominant direction is $i \to \alpha$ with probability of $O(1)$. Furthermore, once the principal eigenvector $\vec{\pi}$ becomes non-degenerate due to the error-induced transition via $\rho_\beta^{(1)}$, all other perturbations $\lesssim O(e)$ that we have neglected add to $\vec{\pi}$ only small changes which vanish in the small-$e$ limit, as can be seen from the fact that the principal eigenvector has a well-defined derivative~\cite{van2007computation}. To summarize, this double-sink example has the following property:
\begin{equation}
\left\{
\begin{array}{ll}
i \to \alpha \text{~and~} \alpha \not\to i & \text{if~} i\in \Omega_\alpha^{(0)}\\
\left\{\begin{array}{l}
i \not\leftrightarrow \alpha\\
i \xrightarrow[1]{} \alpha \text{~and~} \alpha \centernot{\xrightarrow[1]{}} i
\end{array}
\right.
& \text{if~} i \in \Omega_\beta^{(0)},
\end{array}
\right.
\label{eq:absorb}
\end{equation}
where $\Omega_\alpha^{(0)} \cup \Omega_\beta^{(0)}$ equals the whole set of nodes by assumption, and Eq.~\eqref{eq:absorb} guarantees that $\pi_\alpha$ approaches $100\%$ in the limit of $e \to 0^+$. In other words, the original basin $\Omega_\alpha^{(0)}$ has been extended to $\Omega_\alpha^{(1)} \equiv \Omega_\alpha^{(0)} \cup \Omega_\beta^{(0)}$ in the sense of Eq.~\eqref{eq:absorb}, and $\pi_\alpha$ approaches $100\%$ as the new basin of $\alpha$ coincides with the whole set of nodes. Another way to rephrase it is to construct $g_1$ by supplementing $g_0$ with additional links from $\gamma$ to the member nodes of $\rho_\gamma^{(1)}$,
where $\gamma \in \{\alpha, \beta \}$ denotes each sink of $g_0$.
In terms of $g_1$, it can be said as follows: We have $\pi_\alpha \to 100\%$ in the small-$e$ limit because every node $i$ in $g_1$ has a certain integer $k_i \in \{0,1\}$ such that
\begin{equation}
\left\{
\begin{array}{ll}
i \centernot{\xleftrightarrow[k]{}} \alpha    & \text{for~}0 \le k<k_i \\
i \xrightarrow[k]{} \alpha \text{~and~} \alpha \centernot{\xrightarrow[k]{}} i    & \text{for~}k=k_i.
\end{array}
\right.
\label{eq:g1}
\end{equation}

As a more concrete example, let us consider a Markovian system with a set of four nodes, $\{\alpha, \alpha', \alpha'', \beta\}$. The transition matrix is given as follows:
\begin{eqnarray}
W &=&
\begin{pmatrix}
1-e & 1 & 0 & 0\\
e & 0 & 0 & e\\
0 & 0 & 0 & 0\\
0 & 0 & 1 & 1-e
\end{pmatrix}
\begin{array}{c}
\alpha \\
\alpha' \\
\alpha'' \\
\beta
\end{array}\\
&& ~~~~\left. \begin{array}{cccc}
\alpha & \alpha' & \alpha'' & \beta
\end{array} \right.
\label{eq:markov1}
\end{eqnarray}
each of whose columns adds up to one. At $e=0$, the system has two disconnected parts, $\Omega_\alpha^{(0)}=\{\alpha, \alpha'\}$ and $\Omega_\beta^{(0)}=\{\alpha'',\beta\}$. The first part with the largest eigenvalue $\lambda_\alpha(e=0)=1$ has the corresponding left and right eigenvectors, $\vec{y}_\alpha = (1,1,0,0)$ and $\vec{x}_\alpha = (1,0,0,0)^{\intercal}$, respectively. Likewise, the second part with the largest eigenvalue $\lambda_\beta(e=0)=1$ has $\vec{y}_\beta=(0,0,1,1)$ and $\vec{x}_\beta=(0,0,0,1)^{\intercal}$. Note that $\vec{y}_\alpha \cdot \vec{x}_\alpha = \vec{y}_\beta \cdot \vec{x}_\beta = 1$ and $\vec{y}_\alpha \cdot \vec{x}_\beta = \vec{y}_\beta \cdot \vec{x}_\alpha = 0$.
It is a usual practice to calculate eigenvalue perturbation~\cite{morone2016collective}, but
due to the two-fold degeneracy of this problem, we have to diagonalize the following $2 \times 2$ matrix:
\begin{equation}
L =
\begin{pmatrix}
\vec{y}_\alpha \cdot W' \cdot \vec{x}_\alpha &
\vec{y}_\alpha \cdot W' \cdot \vec{x}_\beta \\
\vec{y}_\beta \cdot W' \cdot \vec{x}_\alpha &
\vec{y}_\beta \cdot W' \cdot \vec{x}_\beta
\end{pmatrix}
= \begin{pmatrix}
0 & 1\\
0 & -1
\end{pmatrix} = PDP^{-1}\label{eq:lambda}
\end{equation}
with
\begin{equation}
P =
\begin{pmatrix}
1 & -1\\
0 & 1
\end{pmatrix}
\end{equation}
and
\begin{equation}
D =
\begin{pmatrix}
0 & 0\\
0 & -1
\end{pmatrix},
\end{equation}
where the prime denotes differentiation with respect to $e$. The diagonal elements of $D$ imply that $\lambda_\alpha(e) = 1 + O(e^2)$ and $\lambda_\beta(e) = 1-e + O(e^2)$ so that probability over $\Omega_\beta^{(0)}$ will eventually be absorbed into that over $\Omega_\alpha^{(0)}$ as soon as $e$ becomes positive. If we look at the structures of the left and right eigenvectors, their dot products with $W'$ in Eq.~\eqref{eq:lambda} clearly show that the important point is whether the error-induced transitions from a `sink' go outside its basin.
Although the first sink $\alpha$ survives in this example, the actual stationary distribution, $\vec{\pi} = \left( 1/(1+e), e/(1+e), 0, 0 \right)^{\intercal}$, slightly differs from $\vec{x}_\alpha = (1,0,0,0)^{\intercal}$ because $\alpha \xrightarrow[1]{} \alpha'$ in Eq.~\eqref{eq:markov1}. However, as we have already expected, the difference is insignificant in the sense that $\vec{\pi}$ converges to $\vec{x}_\alpha$ continuously as $e \to 0$.
From $\pi_{\alpha''} / \pi_{\alpha'} = 0$,
we also note that transition $\alpha'' \xrightarrow[1]{} \alpha'$ eventually occurs with probability $e + (1-e)e + (1-e)^2 e + \ldots = 1$ in the long run because of the self-loop at $\beta$. Although it is involved with an error with probability $e \ll 1$, the total probability in the long run may be of $O(1)$, and this is what affects the stationary distribution $\vec{\pi}$.
In addition, as long as the relevant connections are all preserved as described in Eq.~\eqref{eq:g1}, the other error-induced transitions, which may actually be involved in calculating $\vec{\pi}$, do not change the conclusion: As an example, suppose that we add to Eq.~\eqref{eq:markov1} transition from every node to every other with probability $e^2$, represented by
\begin{equation}
\left( \delta W \right)_{ij} = \left\{
\begin{array}{cc}
-3e^2     & \text{if~} i=j \\
e^2     & \text{if~} i\neq j.
\end{array}\right.
\end{equation}
The second example is to add a transition of probability $e$ from $\alpha'$ to $\alpha''$ by setting $W_{\alpha'' \alpha'} = e$ and $W_{\alpha \alpha'} = 1-e$ in Eq.~\eqref{eq:markov1}.
In both of these examples, we can show by direct calculation that the resulting $\vec{\pi}$ keeps converging to $\vec{x}_\alpha$ in the small-$e$ limit. On the other hand, if we add a transition from $\alpha$ to $\alpha''$ with probability $e$, it extends $\rho_\alpha^{(1)}$ beyond $\Omega_\alpha^{(0)} = \{ \alpha, \alpha' \}$, so neither $\alpha$ nor $\beta$ survives alone but they divide up the stationary distribution in the sense that $\pi_\alpha / \pi_\beta \sim O(1)$.

Now, the above procedure can be carried out recursively:
\begin{enumerate}
    \item Construct $g_0$ with a set of nodes, $\mathcal{N}$.
    \begin{itemize}
        \item If $0 \to i$ for a certain node $i$, the strategy is inefficient.
        \item If $\Omega_0^{(0)} \equiv \{i| i \to 0 \text{~and~} 0 \not\to i\}$ equals $\mathcal{N}$, the strategy is efficient.
        \item If $\Omega_0^{(0)}$ is a strict subset of $\mathcal{N}$, the efficiency criterion is undecidable from $g_0$. Go to the next step with $\nu=1$.
    \end{itemize}
    \item Construct $g_\nu$ by adding links from every sink $\gamma$ surviving in $g_{\nu-1}$ to the member nodes of $\rho_\gamma^{(\nu)}$.
    \begin{itemize}
        \item If $0 \xrightarrow[\nu]{} i$ for a node $i$ outside $\Omega_0^{(\nu-1)}$, the strategy is inefficient.
        \item If $\Omega_0^{(\nu)} \equiv \Omega_0^{(\nu-1)} \cup \{i | i \xrightarrow[\nu]{} 0 \text{~and~} 0 \centernot{\xrightarrow[\nu]{}} i \}$ equals $\mathcal{N}$, the strategy is efficient.
        \item If $\Omega_0^{(\nu)}$ is a strict subset of $\mathcal{N}$, the efficiency criterion is undecidable from $g_\nu$. Go to the next step.
    \end{itemize}
    \item Increase $\nu$ by one, and go back to the previous step.
\end{enumerate}
This algorithm always ends with a decision between `efficient' and `inefficient' because the graph becomes strongly connected if we include all possible types of error.
A pseudo-code to judge efficiency is given in Fig.~\ref{code:efficiency}.

\begin{figure}
\begin{lstlisting}[language=Ruby]
def is_efficient(strategy)
  judged = Array(64, false)
  judged[0] = true
  # => judged = [true, false, false, ..., false]

  # initialize g_n = g(S,S)
  gn = construct_g(strategy, strategy)

  until judged.all?
    64.times do |i|
      next if judged[i]
      if gn.reachable(0, i)     # 0->i
        return false            # judged as inefficient
      end
      if gn.reachable(i, 0)     # i->0 && 0!->i
        judged[i] = true
      end
    end
    gn = update_gn(gn)          # g_n <- g_{n+1}
  end
  return true
end

def update_gn(g)
  g_new = g.clone
  sink_sccs = g.sink_strongly_connected_components
  sink_sccs.each do |sink|
    for_each_node_in(sink) do |n|
      # states reachable by an error from n.
      noised_states = [n^1,n^8]
      noised_states.each do |to|
        g_new.add_link(n, to) unless g_new.has_link?(n, to)
      end
    end
  end
  return g_new
end
\end{lstlisting}
\caption{A pseudo-code to judge efficiency of a strategy.}
\label{code:efficiency}
\end{figure}

\begin{figure}
\begin{lstlisting}[language=Ruby]
def is_distinguishable(strategy)
  judged = Array(64, false)
  judged[0] = true
  # => judged = [true, false, false, ..., false]

  # initialize g_n = g(S, AllC)
  gn = construct_g(strategy, AllC)

  until judged.all?
    64.times do |i|
      next if judged[i]
      if gn.reachable(0, i)     # 0->i
        return true             # judged as distinguishable
      end
      if gn.reachable(i, 0)     # i->0 && 0!->i
        judged[i] = true
      end
    end
    gn = update_gn(gn)          # g_n <- g_{n+1}
  end
  return false
end
\end{lstlisting}
\caption{A pseudo-code to judge distinguishability of a strategy. The method `update\_gn' is identical to the one in Fig.~\ref{code:efficiency}.}
\label{code:distinguishability}
\end{figure}

The algorithm to judge the distinguishability criterion is similar to the one for the efficiency criterion as shown in the following.
If $S$ is a distinguishable strategy, the stationary probability of state $\lim_{e\to 0}\pi_{0} < 1$ when players of $S$ and AllC play the game.
This is because an AllC player must have a smaller payoff than that of the co-player except at full cooperation.
Although this was judged by a linear algebraic calculation~\cite{yi2017combination}, we can directly use the topological structure of a graph $g(S,\textrm{AllC})$ just as in the case of the efficiency criterion.
To this end, we only have to construct graphs $g_{\nu}$ from $g_{\nu-1}$, where $g_0$ is $g(S, \textrm{AllC})$ (Fig.~\ref{code:distinguishability}).
%Thus, the following statement is true.
%\begin{theorem}
%Iff $\forall i \in \{1,\dots,63\}$, $\exists n$ s.t. $i\rightarrow 0$ and $0\not\rightarrow i$ in $g_{n}$, strategy $S$ does not satisfy the distinguishability condition.
%\end{theorem}

\subsection*{Exploring the memory-three strategy space}

\begin{figure}
\begin{center}
\includegraphics[width=0.9\textwidth]{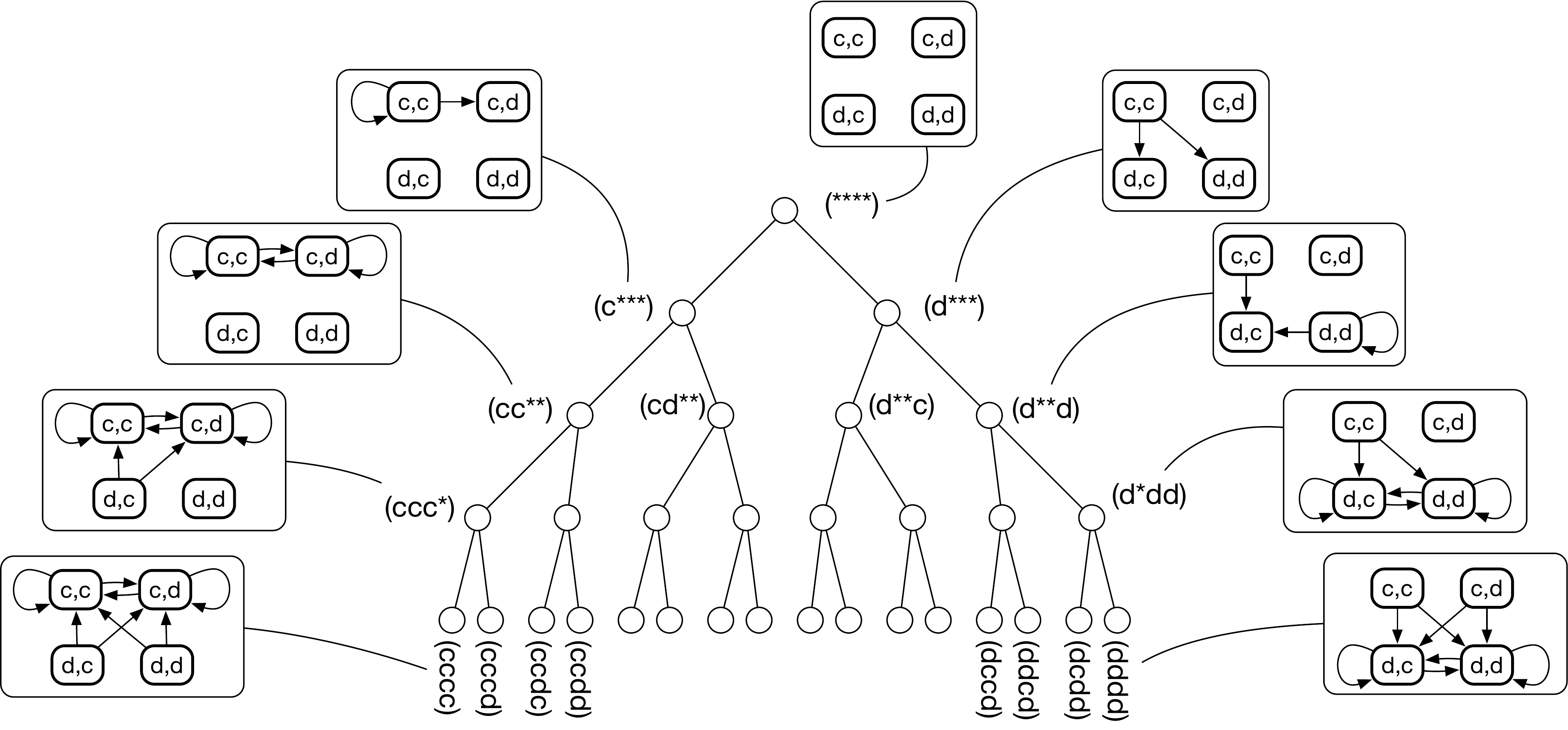}
\end{center}
\caption{
An example of a strategy tree of the memory-$1$ strategy space for the iterated PD game.
A memory-$1$ strategy is represented by a binary string of length $2^{nm} = 4$, each of which corresponds to a leaf node of the strategy tree.
The depth of the strategy tree is $4$.
Starting from the node, which corresponds to the strategy set whose actions are not determined at all,
one of the actions is determined whenever we visit a child node.
Internal nodes correspond to a strategy set. Examples of the transition graphs for strategy sets $g(\mathbf{S},\ast)$ are shown as well.
}
\label{fig:strategy_tree}
\end{figure}

%As mentioned above, the number of candidates in memory-three strategy space is $2^{2^{6}} = 2^{64} \approx 2 \times 10^{19}$.
%This is far beyond the computational ability even using a cutting edge supercomputer hence we need an algorithm to reduce the number of candidates.
The space of memory-$m$ strategies is represented by a complete binary tree of depth $2^{2m}$.
Figure~\ref{fig:strategy_tree} shows a tree of memory-one strategies.
In this representation, a leaf vertex (a vertex having no child vertices) corresponds to a specific strategy whereas an internal vertex (a vertex having child vertices) represents a set of strategies, a part of whose actions remain undetermined.
The root vertex of the tree corresponds to the whole set of strategies in memory-$m$ strategy space.
Traversing the tree from the root to a leaf vertex by one step is equivalent to determining one of the undetermined actions.
Hereafter, this sort of tree is called a strategy tree, and the set of strategies corresponding to an internal vertex is called a strategy set.
The order in which actions are determined may be arbitrary in each subtree.
For example, in Fig.~\ref{fig:strategy_tree}, the root vertex branches into two subtrees depending on what to do at $(c,c)$. If the answer is $c$, we enter the left subtree, and the next question concerns what to do at $(c,d)$. If we choose $d$ at $(c,c)$, on the other hand, we get into the right subtree, and what comes next is the choice at $(d,d)$.

We will begin by finding strategies that satisfy the efficiency and defensibility criteria.
As will be explained below, we focus on necessary conditions for a strategy set to satisfy the efficiency or defensibility criteria: If the necessary conditions are violated at an internal vertex, the whole branch below it may be discarded without further consideration. For this reason, the computational cost crucially depends on in which order the actions are determined along the branches of the tree.

\subsection*{Checking the defensibility criterion for a strategy set}

Let $g(\mathbf{S},\ast)$ be the transition graph for a strategy set $\mathbf{S}$, which is defined as the largest common subgraph of $g(S,\ast)$ for every member strategy $S \in \mathbf{S}$, where $\ast$ is a wildcard character:
If $g(\mathbf{S},\ast)$ contains a negative cycle, $g(S,\ast)$ must also contain it, hence $S$ violates defensibility. Conversely, a necessary condition for $S$ to satisfy the defensibility criterion is the absence of negative cycles in $g(\mathbf{S},\ast)$.
Examples of $g(\mathbf{S},\ast)$ are shown in Fig.~\ref{fig:strategy_tree}.
The transition graph for a strategy set is constructed in the following way:
If an action at one of its nodes (i.e. history profiles) is determined,
two outgoing links are added at the node. They are two because the co-player's choice can be either $c$ or $d$, which leads to a different history profile at the next time step.
If we have not determined the action, the node has no outgoing links.

The existence of negative cycles in $g(\mathbf{S},\ast)$ is judged by the Floyd-Warshall (FW) algorithm~\cite{hougardy2010floyd}.
The FW algorithm finds the minimum distance for every pair of nodes which do not belong to a negative cycle. By distance, we mean the relative payoff difference between the players, so that outgoing links from `positive nodes' $(**d,**c)$ and `negative nodes' $(**c,**d)$ contribute $+1$ and $-1$ to the distance, respectively. The other nodes such as $(**c,**c)$ and $(**d,**d)$ contribute zero and will be called `neutral'.
Specifically, we use the following algorithm:
\begin{enumerate}
    \item Start from the root vertex of the tree. Define $\delta$ as a $64 \times 64$ matrix of the minimum distances for all node pairs. All its elements are formally regarded as $+\infty$ at the root vertex, where no links exist yet.
    \item Move to one of the child vertices, whose corresponding strategy set is denoted by $\mathbf{S}$, by determining an action at node $k$. This corresponds to adding two links to $k$, and we denote these links as $k \to u$ and $k \to v$, respectively.
    \begin{enumerate}
        \item Update the minimum distance between $k$ and an arbitrary node $j$ by calculating $\delta_{kj} = \min\{ \delta_{kj}, \delta_{k u} + \delta_{u j}, \delta_{k v} + \delta_{v j}) \}$.
        \item For the every other pair of nodes $i$ and $j$, update their minimum distance by calculating $\delta_{ij} = \min\{\delta_{ij}, \delta_{ik}+\delta_{kj}\}$.
        \item If the updated matrix $\delta$ has a negative diagonal element, a negative cycle exists in $g(\mathbf{S}, \ast)$. Do not go deeper into this branch. Otherwise, proceed to one of the grandchild vertices recursively as in the depth-first search.
    \end{enumerate}
    \item Check the other child vertex in the same way.
\end{enumerate}
To discard strategies that are not defensible as early as possible, we should begin by checking actions that are likely to form negative cycles:
If a negative node exists with undetermined actions, this should be checked first, by adding outgoing links to the node. For example, in Fig.~\ref{fig:strategy_tree}, we can say that a strategy violates defensibility if it prescribes $c$ at its unique negative node $(c,d)$ because such prescription forms a negative cycle of $(c,d) \to (c,d) \to \ldots$. Provided that $d$ is the correct action at $(c,d)$, let us proceed to one of the subsequent nodes, $(d,d)$. One must choose $d$ here: Otherwise, we will see a negative cycle $(d,d) \to (c,d) \to (d,d) \to \ldots$. After determining these two actions, the strategy set $\mathbf{S}$ can be written as $(*d,*d)$, and it is no longer possible to form a negative cycle at this point: To revisit the negative node $(c,d)$ to complete a cycle, one must go through the positive node $(d,c)$. We can say that all the possible cycles from the negative node have been \emph{neutralized} in this strategy set $\mathbf{S} = (*d,*d)$. With just two steps, this procedure gives the list of memory-one strategies that satisfy denfensibility.
In general, we will use the following procedure:
\begin{enumerate}
\item Determine actions at all the negative nodes, among which $d$ must be chosen at $(\underbrace{c\cdots c}_{m}, \underbrace{d\cdots d}_{m})$ for obvious reason.
\item If we have not determined action at a node, we will call the node `susceptible'. Let $K$ be the set of susceptible nodes linked from the negative nodes.
\item For each $k \in K$, compute $y_k \equiv \min_{j \in G} \delta_{jk}$, where $G$ is the set of negative nodes.
\begin{itemize}
    \item If $k$ is a positive node with $y_k = -1$, remove it from $K$ because this path is neutralized.
    \item Otherwise, add two outgoing links to $k$ by determining an action. Replace $k$ in $K$ by its subsequent susceptible nodes.
\end{itemize}
\item Repeat Step 3 until a negative cycle is found or $K$ becomes empty. In the latter case, all the strategies in the remaining strategy set do not have a negative cycle.
\end{enumerate}

%a node with undetermined actions, say, $j$, with $\min_{i \in K} \delta_{ij} < 0$, where $K$ means the set of negative nodes (Fig.~\ref{fig:negative_dangling_nodes}).
%If $g(\mathbf{S},\ast)$ has none of them, the graph cannot contain a negative cycle, and all the member strategies of $\mathbf{S}$ satisfy defensibility.

%\begin{figure}
%\begin{center}
%\includegraphics[width=0.8\textwidth]{negative_dangling_nodes.pdf}
%\end{center}
%\caption{
%An example of negative unfixed nodes and negative dangling nodes.
%Red (blue) nodes denote negative (positive) nodes while the white ones are neutral, %i.e., (**c,**c) or (**d,**d).
%Dotted nodes represent nodes at which actions are not fixed.
%}
%\label{fig:negative_dangling_nodes}
%\end{figure}

\subsection*{Checking the efficiency criterion for a strategy set}

\begin{figure}
\begin{center}
\includegraphics[width=0.9\textwidth]{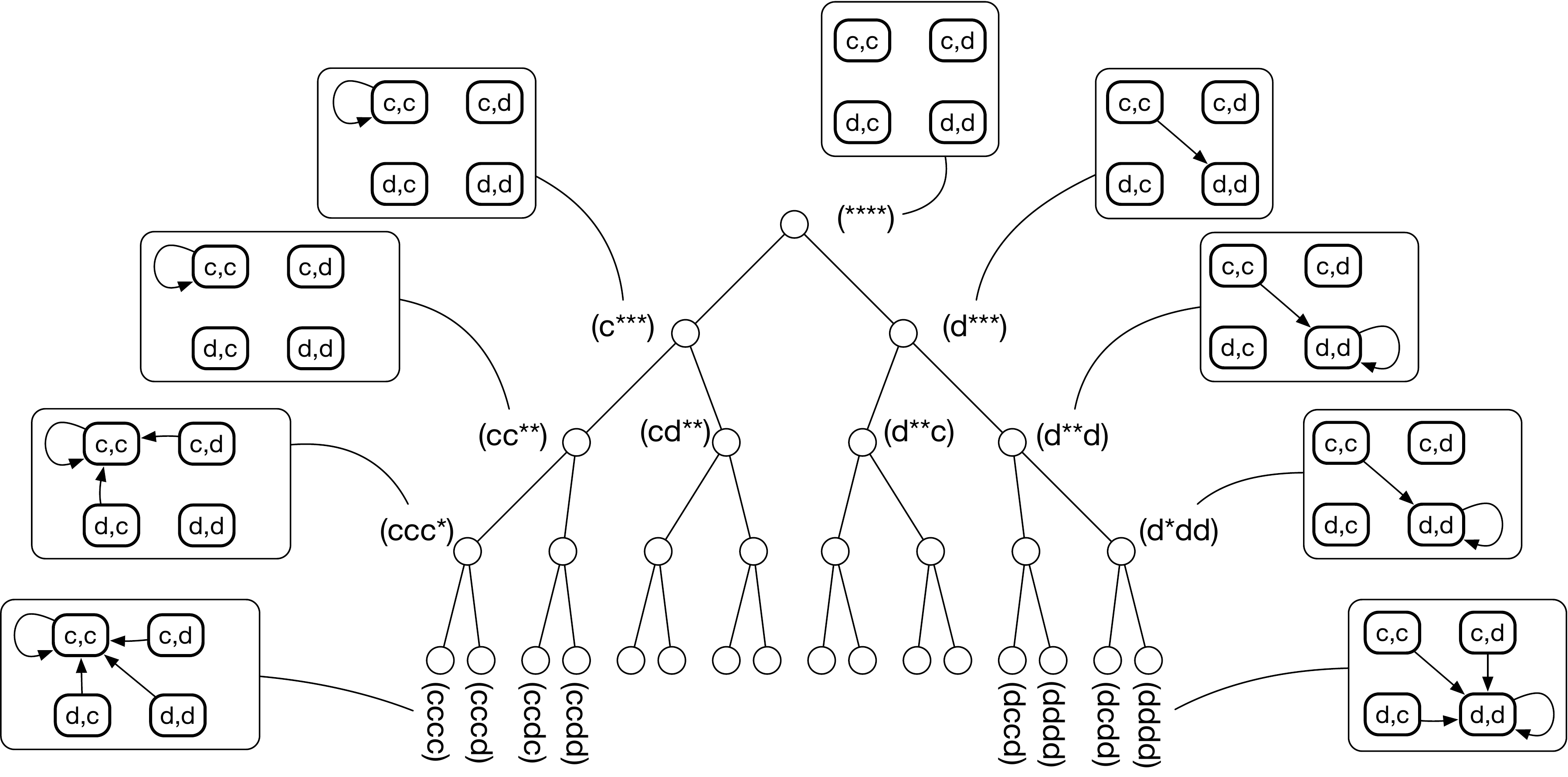}
\end{center}
\caption{
An example of a strategy tree of the memory-one strategy space and $g(\mathbf{S}, \mathbf{S})$, the largest common part of $g(S,S)$ for every member strategy $S \in \mathbf{S}$.
Note that links may or may not be added when we go down the tree by one step.
Compare this with Fig.~\ref{fig:strategy_tree} where we have depicted $g(\mathbf{S},\ast)$.
}
\label{fig:strategy_tree_gSS}
\end{figure}

Similarly, a necessary condition exists for a strategy set to satisfy the efficiency criterion.
First, an efficient strategy needs to recover mutual cooperation against one-bit error at least.
The transition from state $8$ $(ccd,ccc)$ and state $1$ $(ccc,ccd)$ must eventually reach state $0$ in $g(S,S)$. Otherwise, it cannot be efficient.
This judgement is useful for a strategy set as well:
We construct a graph $g(\mathbf{S},\mathbf{S})$ for a strategy set $\mathbf{S}$, which is defined as the largest common subgraph of $g(S,S)$ for every member strategy $S \in \mathbf{S}$ (Fig.~\ref{fig:strategy_tree_gSS}).
For example, if we trace the transition from state $1$ or $8$ in this graph and find a cycle other than $0$, all the strategies in $\mathbf{S}$ cannot be efficient, and thus it is not necessary to go further than this strategy set.
%To reject inefficient strategies as early as possible,
%we should determine actions in a way that generates paths from states $1$ and $8$ first.

The above method checks whether the mutual cooperation is tolerant against one-bit error, which is a necessary condition for efficiency.
To assure the efficiency of a strategy set, we need to take into account higher-order terms of $e$ as well.
The key observation is that only SCC's can occupy finite stationary probability, which is an essential object in judging efficiency.
By extending the graph-theoretic method in Fig.~\ref{code:efficiency}, we have developed a method to judge efficiency for a strategy set as follows:
%First, we construct a graph $\hat{g}(\mathbf{S},\mathbf{S})$ for a strategy set $\mathbf{S}$, which is defined as the \emph{union} of $g(S,S)$ for $S \in \mathbf{S}$.
%Then, we obtain strongly connected components in $\hat{g}(\mathbf{S},\mathbf{S})$, and we denote as $\mathcal{\hat{C}}$ the set of nodes constituting these strongly connected components.
%Since $g(\mathbf{S},\mathbf{S})$ is a subgraph of $\hat{g}(\mathbf{S},\mathbf{S})$, the nodes constituting the strongly connected components in $g(\mathbf{S},\mathbf{S})$, denoted as $\mathcal{C}$, is a subset of $\mathcal{\hat{C}}$, i.e., $\mathcal{C} \subset \mathcal{\hat{C}}$.
%If any of the actions at $\mathcal{\hat{C}}$ is unfixed, it is not possible to determine the efficiency of $\mathbf{S}$.
%In this case, we traverse the strategy tree and apply the same procedure recursively to its child strategy sets.
%If the actions at $\mathcal{\hat{C}}$ are all fixed, it is sure that the actions at $\mathcal{C}$ are fixed as well.
%In other words, the nodes that are potentially be a part of the strongly connected component in $g(S,S)$ are all fixed and the strongly connected components $g(\mathbf{S},\mathbf{S})$ are identical to those in $g(S,S)$ for all $S \in \mathbf{S}$.
Let $\mathcal{C}$ denote the set of nodes constituting the SCC's in $g(\mathbf{S},\mathbf{S})$. We go down the tree until every node in $\mathcal{C}$ has a prescribed action. Then,
the SCC's of $g(\mathbf{S},\mathbf{S})$ will be identical to those in $g(S,S)$ for every $S \in \mathbf{S}$.
When this is the case, we use the following algorithm to test the efficiency of $\mathbf{S}$:
\begin{enumerate}
    \item Calculate $\mathcal{C}$ for $g(\mathbf{S},\mathbf{S})$.
    \item Construct $g_0 = g(\mathbf{S},\mathbf{S})$.
    \begin{itemize}
        \item If $0 \to i$ for a certain node $i$, the strategy is inefficient.
        \item If $\Omega_0^{(0)} \equiv \{i \in \mathcal{C}| i \to 0 \text{~and~} 0 \not\to i\}$ equals $\mathcal{C}$, the strategy is efficient.
        \item If $\Omega_0^{(0)}$ is a strict subset of $\mathcal{C}$, the efficiency criterion is undecidable from $g_0$. Go to the next step with $\nu=1$.
    \end{itemize}
    \item Construct $g_\nu$ by adding links from every sink $\gamma$ surviving in $g_{\nu-1}$ to the member nodes of $\rho_\gamma^{(\nu)}$.
    If $g_\nu$ has an unfixed node which is reachable from $\mathcal{C}$, fix the action at the nodes and then apply the same sequence recursively to its child strategy sets.
    \begin{itemize}
        \item If $0 \xrightarrow[\nu]{} i$ for a node $i$ outside $\Omega_0^{(\nu-1)}$, the strategy is inefficient.
        \item If $\Omega_0^{(\nu)} \equiv \Omega_0^{(\nu-1)} \cup \{i \in \mathcal{C} | i \xrightarrow[\nu]{} 0 \text{~and~} 0 \centernot{\xrightarrow[\nu]{}} i \}$ equals $\mathcal{C}$, the strategy is efficient.
        \item If $\Omega_0^{(\nu)}$ is a strict subset of $\mathcal{C}$, the efficiency criterion is undecidable from $g_\nu$. Go to the next step.
    \end{itemize}
    \item Increase $\nu$ by one, and go back to the previous step.
\end{enumerate}
This algorithm always ends with a decision between `efficient' and `inefficient' as in the case of the algorithm for a single strategy.

\subsection*{Overall workflow}

Another tip to reduce the number of strategies is checking the efficiency and defensibility criteria simultaneously.
While the number of strategies satisfying either one of the criteria is
enormous, the number is significantly reduced by checking for the efficiency and
the defensibility criteria simultaneously because these two criteria require
apparently contradictory behaviours (see Fig.~1 in the main text).
The overall workflow is thus organized as follows:
\begin{enumerate}
    \item Traverse the strategy tree with checking the defensibility criterion. The traversal goes down to a certain depth $D_d$.
    \item Traverse the strategy tree with checking the efficiency criterion. The traversal goes down to $D_e$.
    \item Repeat the above two steps with changing parameters.
\end{enumerate}
If $D_d$ or $D_e$ is large, the number of strategy sets increases exponentially.
Switching between these two steps with small depths is important
to carry out the calculation in practice.
We have tested various values of $D_d$ and $D_e$, and found that it does not change the resulting number of successful strategies. We have also checked defensibility and efficiency
of strategies that are randomly chosen from our calculation result. In short, we can say that the algorithm works as intended.

\section*{Supplementary Examples}
\label{appendix:example}

To understand the mechanisms for successfulness in detail, we will give two examples of successful strategies, denoted by ES1 and ES2, respectively. The former is taken from the strategies having the shortest recovery path and the latter is from those with a longer recovery path.

\begin{table}[htb]
\begin{center}
\caption{
Action table of the first example of memory-three successful strategies, which is denoted as ES1.
% cdddcdcdcdcddccddcdccdcdccdcdccdcdcdcccdcdcddcdddcdccdddccddcddd
}\label{tab:first_example}
\begin{tabular}{|c|cccccccc|}
\hline
      & \multicolumn{8}{c|}{$B_{t-3}B_{t-2}B_{t-1}$} \\
$A_{t-3}A_{t-2}A_{t-1}$ & $ccc$ & $ccd$ & $cdc$ & $cdd$ & $dcc$ & $dcd$ & $ddc$ & $ddd$\\
\hline
$ccc$ & $c$ & $d$ & $d$ & $d$ & $c$ & $d$ & $c$ & $d$ \\
$ccd$ & $c$ & $d$ & $c$ & $d$ & $d$ & $c$ & $c$ & $d$ \\
$cdc$ & $d$ & $c$ & $d$ & $c$ & $c$ & $d$ & $c$ & $d$ \\
$cdd$ & $c$ & $c$ & $d$ & $c$ & $d$ & $c$ & $c$ & $d$ \\
$dcc$ & $c$ & $d$ & $c$ & $d$ & $c$ & $c$ & $c$ & $d$ \\
$dcd$ & $c$ & $d$ & $c$ & $d$ & $d$ & $c$ & $d$ & $d$ \\
$ddc$ & $d$ & $c$ & $d$ & $c$ & $c$ & $d$ & $d$ & $d$ \\
$ddd$ & $c$ & $c$ & $d$ & $d$ & $c$ & $d$ & $d$ & $d$ \\
\hline
\end{tabular}
\end{center}
\end{table}

The first example of memory-three successful strategies is defined by Table~\ref{tab:first_example} and denoted as ES1.
The behaviour of this strategy is distinct from TFT-ATFT in several respects:
Let us look at the mechanism to stabilize the cooperation.
When the players using this strategy, the mutual cooperation is recovered from a
one-bit error in two steps, as we see from the first entry of Table~1 in the
main text.

\begin{figure}
\begin{center}
\includegraphics[width=0.9\textwidth]{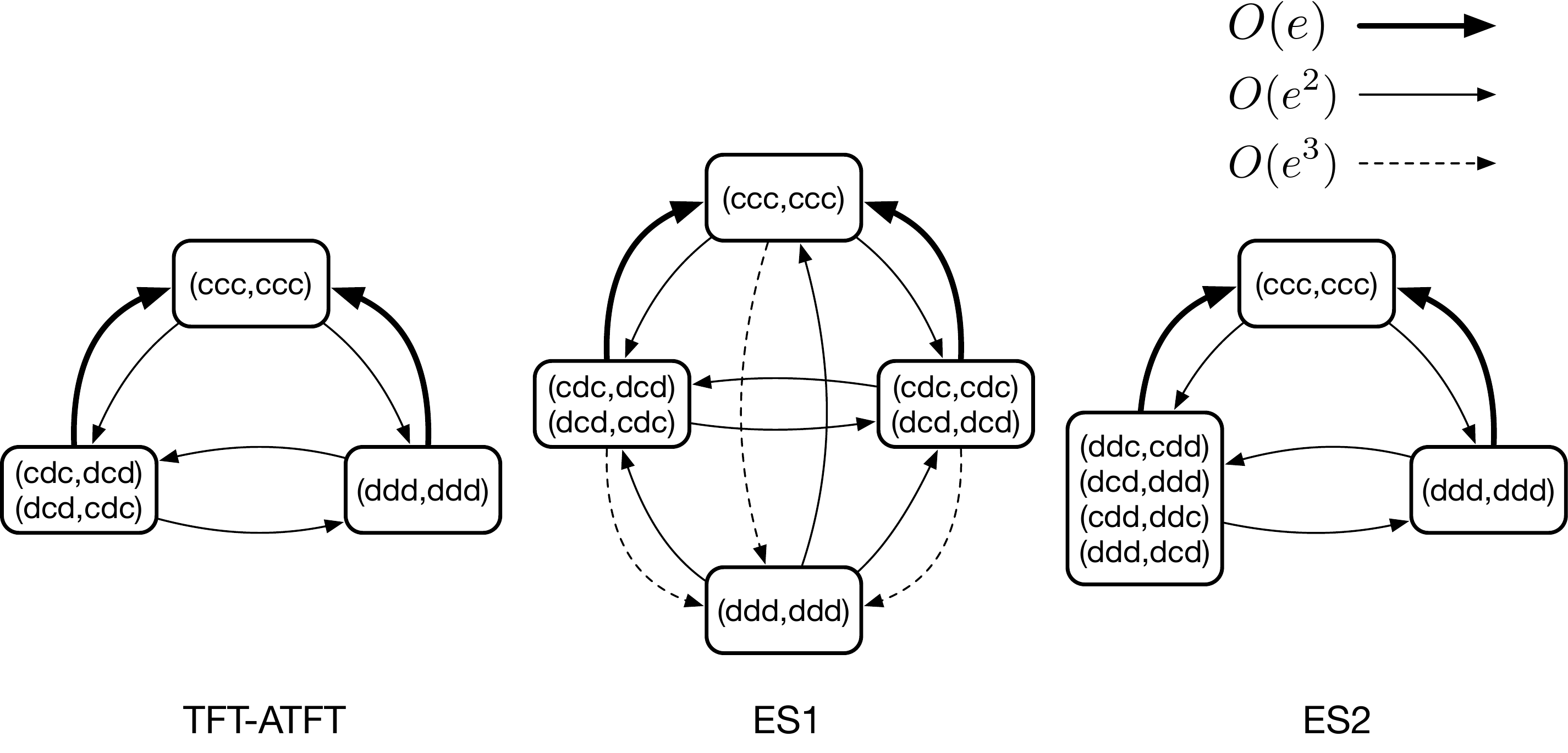}
\end{center}
\caption{
Transitions between strongly connected components (or cycles) in $g(S,S)$ for $S=$TFT-ATFT, ES1, and ES2.
The thick, thin, and dashed arrows indicate transitions with probabilities $O(e)$, $O(e^2)$, and $O(e^3)$, respectively.
}
\label{fig:transition_probs}
\end{figure}

Although mutual cooperation is robust against a one-bit error, it does not assure that the strategy meets the efficiency criterion.
This is because mutual defection is also robust against a one-bit error:
When an implementation error occurs at state $63 = (ddd,ddd)$, the state eventually returns to mutual defection as
\begin{equation}
(ddd,ddc) \rightarrow (ddd,dcd) \rightarrow (ddd,cdd) \rightarrow (ddd,ddd).
\end{equation}
This error robustness of mutual defection is not found in TFT-ATFT.
To see how the efficiency criterion is satisfied, we need to look at higher-order transitions mediated by more than one errors.
Figure~\ref{fig:transition_probs} shows the transition between the cycles in $g(\text{ES1}, \text{ES1})$.
The graph has four cycles: (i) mutual cooperation $(ccc,ccc)$, (ii) mutual defection $(ddd,ddd)$,
(iii) TFT retaliation $(cdc,dcd) \leftrightarrow (dcd,cdc)$, and (iv) synchronous repetition of cooperation and defection $(cdc,cdc) \leftrightarrow (dcd,dcd)$.
The transition from mutual cooperation to mutual defection occurs with $O(e^3)$ while the transition for the opposite direction happens with $O(e^2)$.
In other words, the net probability flow is towards mutual cooperation, whereby the efficiency criterion is fulfilled.
This is distinct from the case for TFT-ATFT, which does not exhibit (iv).
The transitions between these cycles for TFT-ATFT are also drawn in Fig.~\ref{fig:transition_probs}:
Because the transition from (ii) or (iii) to (i) occurs with $O(e)$, it is sufficient to make the mutual cooperation tolerant against one-bit error to assure efficiency.

\begin{figure}
\begin{center}
\includegraphics[width=0.9\textwidth]{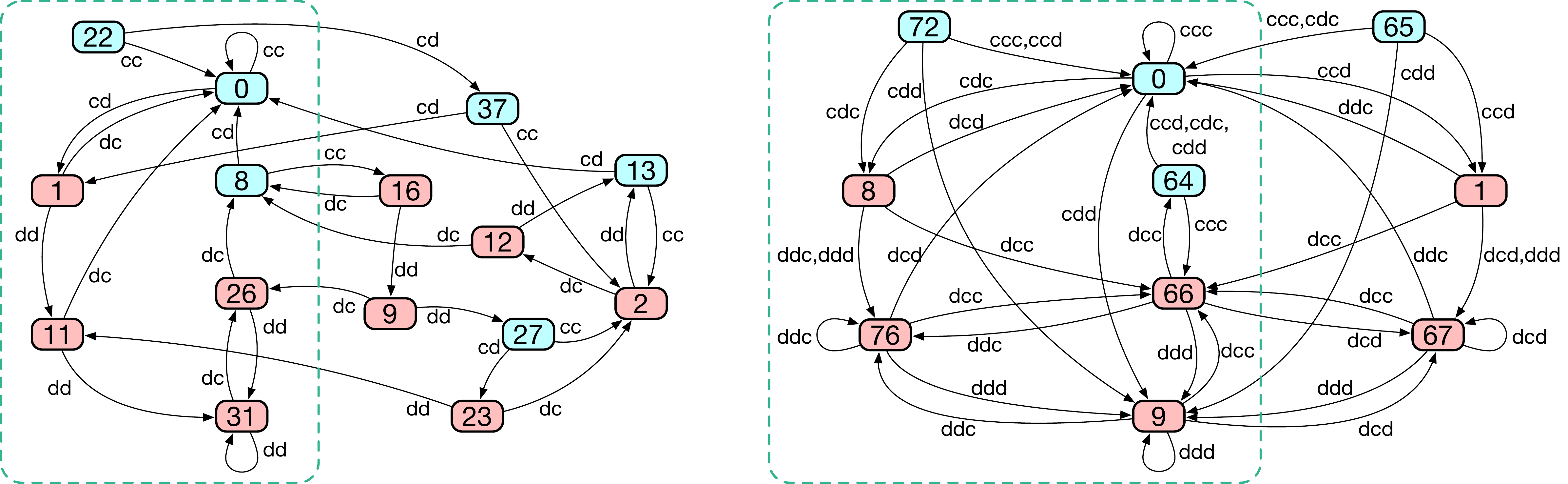}
\end{center}
\caption{
Automaton representations of ES1 (left) and a successful strategy for the three-person PG game (right).
Colours of nodes represent prescribed actions at the corresponding history profiles: Cooperation (defection) is prescribed at blue (orange) nodes.
The label on each edge means the actions taken at the last time step, $(A_{t-1}B_{t-1})$ (or $(A_{t-1}B_{t-1}C_{t-1})$ in case of the three-person game).
Note the similarity between the two strategies as indicated by the dashed boxes.
}
\label{fig:es1_automaton}
\end{figure}

It is also instructive to convert a strategy defined by an action table to an automaton having the minimal number of states~\cite{murase2019automata}.
Figure~\ref{fig:es1_automaton} shows an automaton derived from ES1, which has $15$ internal states.
Compared to the automaton for TFT-ATFT~\cite{murase2019automata}, it has a greater number of states with a very different graph structure.
Actually, it bears more similarity to that of a successful strategy for the three-person PG game (see the dashed boxes in Fig.~\ref{fig:es1_automaton}), and it is not a coincidence:
Both of these two strategies have $m=3$, and mutual cooperation is robust against two-bit errors, whereas the mutual defection is robust against one-bit error~\cite{murase2018seven}.

\begin{figure}
\begin{center}
\includegraphics[width=0.5\textwidth]{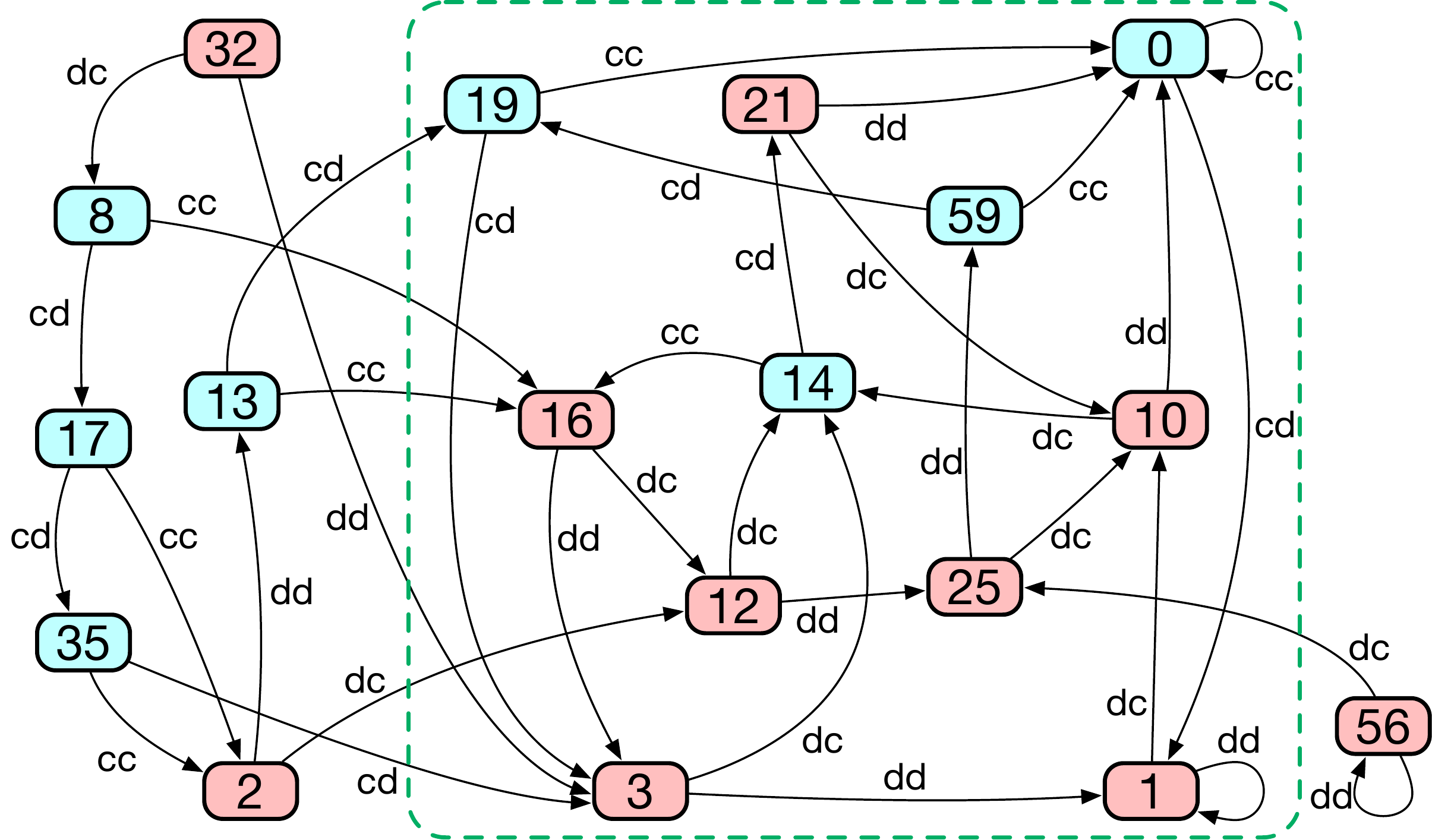}
\end{center}
\caption{
Automaton representation of the second example strategy (ES2).
The dashed rectangle indicates the strongly connected components of the strategy.
}
\label{fig:ES2_automaton}
\end{figure}

Another example strategy (ES2) is defined by Table~\ref{tab:ES2}, whose
automaton representation is given in Fig.~\ref{fig:ES2_automaton}.
Obviously, this is not a variant of the TFT-ATFT strategy, and the path to recover mutual cooperation is much longer than that of ES1 or TFT-ATFT:
\begin{multline}
(1,8) \rightarrow (10,17) \rightarrow (14,35) \rightarrow (16,2) \rightarrow (3,13) \rightarrow (14,19) \rightarrow \\
(16,0) \rightarrow (12,1) \rightarrow (25,1) \rightarrow (59,1) \rightarrow (19,10) \rightarrow (3,14) \rightarrow \\
(14,21) \rightarrow (21,10) \rightarrow (0,0).
\end{multline}
Efficiency of ES2 is explained by Fig.~\ref{fig:transition_probs} which depicts
strongly connected components in $g(\text{ES2},\text{ES2})$ and transition among them.
Defensibility is verified by Fig.~\ref{fig:ES2_automaton} because it has no negative cycle.
The distinguishability criterion is also satisfied because of the cycle $14 \rightarrow 16 \rightarrow 12$, with which ES2 can repeatedly exploit an AllC player.

So far, we have focused on distinguishability only against AllC players.
Generalizing this idea, we can think of a strategy that can distinguish not only AllC but also a broader class of non-defensible strategies.
Let us take WSLS as an example of non-defensible strategies.
When TFT-ATFT meets WSLS, they do not achieve full cooperation, but
they get the same long-term payoff when $e \to 0$, indicating that TFT-ATFT cannot distinguish a WSLS player.
On the other hand, ES1 is able to distinguish a WSLS player in the sense that
the long-term payoff of ES1 is strictly higher than that of WSLS,
and ES1 satisfies the extended distinguishability criterion.
Finally, when ES2 plays against WSLS, they form full cooperation.
Thus, these three successful strategies, TFT-ATFT, ES1, and ES2, show different behaviours against WSLS.

\begin{table}[htb]
\begin{center}
\caption{
An action table of the second example of memory-three successful strategies (ES2).
%cdddcddd cddddccd dcdcddcd cdcdccdddddccdcddddcddddddcdddccdddcccdd
}\label{tab:ES2}
\begin{tabular}{|c|cccccccc|}
\hline
      & \multicolumn{8}{c|}{$B_{t-3}B_{t-2}B_{t-1}$} \\
$A_{t-3}A_{t-2}A_{t-1}$ & $ccc$ & $ccd$ & $cdc$ & $cdd$ & $dcc$ & $dcd$ & $ddc$ & $ddd$\\
\hline
$ccc$ & $c$ & $d$ & $d$ & $d$ & $c$ & $d$ & $d$ & $d$ \\
$ccd$ & $c$ & $d$ & $d$ & $d$ & $d$ & $c$ & $c$ & $d$ \\
$cdc$ & $d$ & $c$ & $d$ & $c$ & $d$ & $d$ & $c$ & $d$ \\
$cdd$ & $c$ & $d$ & $c$ & $d$ & $c$ & $c$ & $d$ & $d$ \\
$dcc$ & $d$ & $d$ & $d$ & $c$ & $c$ & $d$ & $c$ & $d$ \\
$dcd$ & $d$ & $d$ & $d$ & $c$ & $d$ & $d$ & $d$ & $d$ \\
$ddc$ & $d$ & $d$ & $c$ & $d$ & $d$ & $d$ & $c$ & $c$ \\
$ddd$ & $d$ & $d$ & $d$ & $c$ & $c$ & $c$ & $d$ & $d$ \\
\hline
\end{tabular}
\end{center}
\end{table}

When ES1 and ES2 play the game,
their long-term payoffs are identical because of the defensibility criterion, but their cooperation probability is below 100\%.
This is because their recovery mechanisms from implementation error are different.
Therefore, we can conclude that different types of successful strategies do not always achieve full cooperation although each of them meets the efficiency criterion.
We have already seen many types of successful strategies in memory-three strategy space.
To achieve full cooperation, players need not only adopt successful strategies but also select the same type of successful strategies. The problem thus boils down to the coordination game.
In the memory-two strategy space, the situation is different because every successful variant of TFT-ATFT achieves full cooperation with every other.

%\bibliographystyle{elsarticle-harv}
%\bibliography{ir}
%\end{document}

\end{document}